\documentclass[review]{elsarticle}

\usepackage{hyperref}
\usepackage[T1]{fontenc}
\usepackage{graphicx}
\usepackage{siunitx}
\usepackage{booktabs}
\usepackage{xcolor}
\usepackage{float}
\usepackage[normal,raggedright,scriptsize]{subfigure}
\usepackage[normalem]{ulem}

\usepackage{lineno,hyperref}
\modulolinenumbers[5]

\journal{Journal of Magnetism and Magnetic Materials}

\begin{document}

\begin{frontmatter}

\title{Magnetic stiffness calculation for the corresponding  force between two current-carrying circular filaments arbitrarily oriented in the space}


\author[mymainaddress,mysecondaryaddress]{Kirill Poletkin\corref{mycorrespondingauthor}}
\ead{k.poletkin@innopolis.ru}
\cortext[mycorrespondingauthor]{Corresponding author}

\author[Babicaddress]{Slobodan Babic\corref{correspondingauthor}}
\ead{slobobob@yahoo.com}
\cortext[correspondingauthor]{Corresponding author}

\address[mymainaddress]{Innopolis University, 1, Universitetskaya Str., Innopolis, 420500, Russia}
\address[mysecondaryaddress]{The Institute of Microstructure Technology, Karlsruhe Institute of Technology,
Hermann-von-Helmholtz-Platz 1, 76344 Eggenstein-Leopoldshafen, Germany}
\address[Babicaddress]{53 Berlioz 101, H3E 1N2, Montr\'{e}al, Qu\'{e}bec, Canada}

\begin{abstract}
In this article,  sets of analytical formulas for calculation of nine components of magnetic stiffness of corresponding force arising between two current-carrying circular filaments arbitrarily oriented in the space are derived by using Babic's method and the method of mutual inductance (Kalantarov-Zeitlin’s method).  Formulas are presented  through integral
expressions, whose kernel function is expressed in terms of the elliptic integrals of the first and second kinds. Also, we obtained an additional set of  expressions for calculation of components of  magnetic stiffness by means of differentiation of Grover’s formula of the mutual inductance between two circular
filaments with respect to appropriate coordinates. The derived sets of formulas were mutually validated and results of calculation of components of magnetic stiffness agree well to each other.
\end{abstract}


\end{frontmatter}

\linenumbers

\section{Introduction}

Analytical and semi-analytical methods in   the calculation of self- and mutual-inductances of conducting elements of electrical circuits and  magnetic force interactions between these elements have become powerful mathematical instruments in development of power transfer, wireless communication,  sensing and actuation and have been applied in a broad fields of science, including electrical and electronic
engineering, medicine, physics, nuclear magnetic resonance, mechatronics and
robotics, to designate the most prominent. Although, a number of efficient numerical methodes implemented in the commercially developed software are available,  analytical and semi-analytical methods allow to obtaining the result of calculations in the form of a
final formula with a finite number of input parameters, which when applicable
may significantly reduce computation effort. Providing the direct access to  a calculational formula for a user in such methods facilitate mathematical
analysis of obtained  results of calculation  and
opens an opportunity for their further  optimization.

Analytical methods applied to the calculation of mutual inductance between
two circular filaments and arising magnetic force, magnetic torque and corresponding magnetic stiffness when such the filament system carries  electric currents is a prime example.
These methods have proved their efficiency and  have been successfully employed in an increasing number of applications, including electromagnetic levitation \cite{OkressWroughtonComenetzEtAl1952,Narukullapati2021}, superconducting levitation \cite{Paredes2021}, 
calculation of mutual inductance between  thick coils \cite{Ravaud2010}, {magnetic force and torque calculation between circular coils \cite{Babic2008,Babic2011,Babic2012}
, wireless power transfer \cite{JowGhovanloo2007,SuLiuHui2009,ChuAvestruz2017}, electromagnetic actuation \cite{ShiriShoulaie2009,RavaudLemarquandLemarquand2009,Obata2013}, 
micro-machined contactless inductive suspensions 
\cite{Poletkin2014a,Lu2014,PoletkinLuWallrabeEtAl2017b} 
and hybrid contactless suspensions \cite{Poletkin2012,PoletkinKorvink2018, Poletkin2020, Poletkin2021}, biomedical applications \cite{TheodoulidisDitchburn2007,SawanHashemiSehilEtAl2009}, topology optimization \cite{KuznetsovGuest2017}, nuclear magnetic resonance \cite{D.I.B.2002,SpenglerWhileMeissnerEtAl2017}, indoor positioning systems \cite{AngelisPaskuAngelisEtAl2015}, navigation sensors \cite{WuJeonMoonEtAl2016}, non-contact gap measurement sensors \cite{Jiao2021},  wireless power transfer systems \cite{Zhang2021,Chu2021}, magneto-inductive wireless communications \cite{Gulbahar2017} and others.

In the present article, the set of formulas for calculation of nine components of magnetic stiffness of corresponding force arising between two current-carrying circular filaments arbitrarily oriented in the space are derived by using two methods, namely, Babic's method and the method of mutual inductance. In the first one, the components of the magnetic field at an arbitrary point of the secondary circular filament  generated by the primary coil carrying eletric current are calculated and then after taking the first derivatives of these components with respect to the appropriate coordinates, the set of analytical formulas for calculation of magnetic stiffness appeared in the integral form whose kernel function is expressed in terms of the elliptic integrals of the first and second kinds is derived.  In the second method, the calculation of components of magnetic stiffness of the corresponding magnetic force is performed by
means of finding the second derivatives of the function of mutual inductance between two circular filaments recieved by using Kalantarov-Zeitlin’s method  \cite{Poletkin2019} with respect to the appropriated coordinates.

The article is organized in the following way. In section 2 of the paper, Babic's method and its basic expressions are introduced and the set of analytical formulas for calculation of nine components of magnetic stiffness is derived based on this method. In section 3, the method of mutual inductance is presented. The section includes the preliminary discussion, where the set of coordinate frames necessary for determining the position of the secondary coil with respect to the primary one by using Grover's angles is given.  Also, the relationship between constants of the inclined plane equation employing to define the angular misaliment of the secondary circular filament with respect to the primary one in Babic's approach and Grover's angles is shown.
In section 4, sets of analytical formulas for calculation of nine components of magnetic stiffness recieved by means of the introduced two approaches  are mutually verified via a number of designed examples.  In section 5, conclusions about obtained results are discussed. In the appendix, in addition to Babic's method and the method of mutual inductance, Grover's method is introduced and a set of analytical formulas for calculation of magnetic stiffness based on Grover's method  is obtained.

\section{Babic's method (BM)}

\begin{figure}[!t]
  \centering
  \includegraphics[width=3.5in]{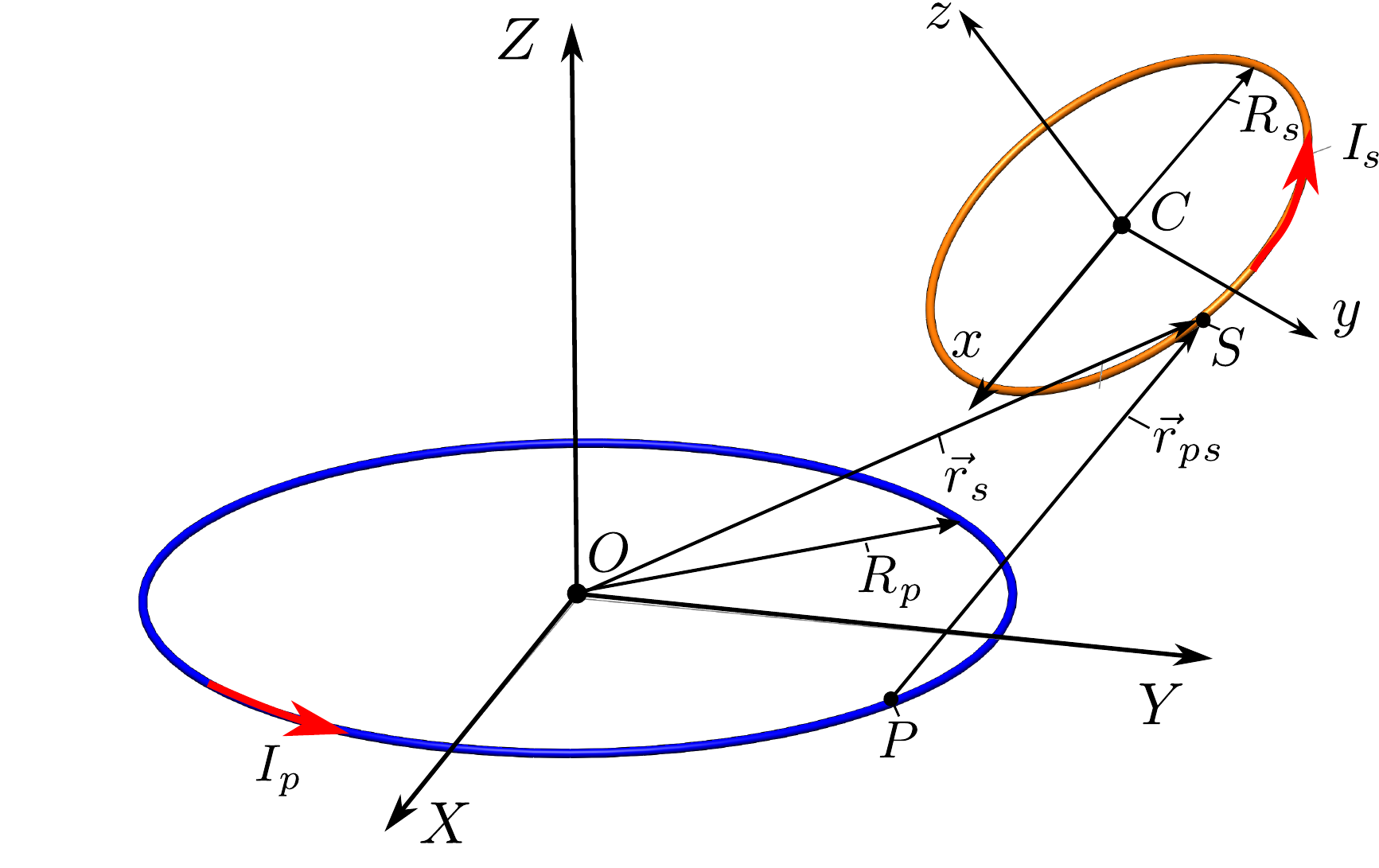}
  \caption{General scheme of arbitrarily positioning two  current-carrying  circular filaments with respect to each other.   }\label{fig:scheme}
\end{figure}

Let us take into consideration two current-carrying circular filaments as showed in Fig. \ref{fig:scheme}, where the center of the larger loop (primary coil) of the radius $R_p$ is placed at the plane $XOY$ whose center is $O$ $(0,0,0)$. The smaller circular loop (secondary coil) of the radius $R_s$ is placed in an inclined plane whose general equation is,
\begin{equation}\label{eq:inclined plane}
{  \lambda \equiv ax+by+cz+d=0,}
\end{equation}
where $a$, $b$, $c$  and $d$ are the components of the normal $\vec{N}$ on the inclined plane in the center of the secondary circular segment $C$ $(x_c, y_c, z_c)$.

\subsection{Basic expressions }
 The segments are with the currents $I_p$ and $I_s$, respectively. For circular filaments (see Fig. \ref{fig:scheme}) we define, \cite{BabicSiroisAkyelEtAl2010,Babic2012a,Babic2012b,Babic2021}:

 \begin{description}
   \item[1)] Since, the primary circular filament  is placed in the plane $XOY$ $(Z = 0)$ with the center at $O$ $(0, 0, 0)$. Hence, an arbitrary point $P$ $(x_p, y_p, z_p)$ of this filament has the following parametric   coordinates (see Fig.\ref{fig:scheme}):
       \begin{equation}\label{eq:parametric coord of PC}
         x_p=R_p\cos\phi, \;y_p=R_p\sin\phi, \;z_p=0,\; \phi\in[0,2\pi].
       \end{equation}
   \item[2)] The differential of the primary circular filament is given by
   \begin{equation}\label{eq:differential coord of PC}
         d\vec{l}_p=R_p\{-\sin\phi, \cos\phi, 0\}d\phi, \; \phi\in[0,2\pi].
       \end{equation}
   \item[3)] The secondary circular filament of radius $R_s$ is placed in the inclined plane (\ref{eq:inclined plane}) with the   center at $C$ $(x_c, y_c, z_c)$. The unit vector $\vec{N}$ (the unit vector of the $z$-axis) at the point $C$, which is the center of the secondary circular filament, laying in the plane $\lambda$ is defined by
   \begin{equation}\label{eq:unit vector N}
         \vec{N}=\left\{\frac{a}{L}, \frac{b}{L}, \frac{c}{L}\right\}, \; L=\sqrt{a^2+b^2+c^2}.
       \end{equation}
   \item[4)] The unit vector between two points C and S they are placed in the plane (\ref{eq:inclined plane}) is
   \begin{equation}\label{eq:unit vector u}
         \vec{u}=\left\{u_x, u_y, u_z\right\}=\left\{-\frac{ab}{Ll}, \frac{l}{L}, -\frac{cb}{Ll}\right\}, \; l=\sqrt{a^2+c^2}.
       \end{equation}
      \item[5)] We define the unite vector $\vec{v}$ lying in the plane (\ref{eq:inclined plane}) and mutually perpendicular on the unit vectors $\vec{N}$ and $\vec{u}$ as the cross-product as follows
   \begin{equation}\label{eq:unit vector v}
         \vec{v}=\vec{N}\times\vec{u}=\left\{v_x, v_y, v_z\right\}=\left\{-\frac{c}{l}, 0, \frac{a}{l}\right\}.
       \end{equation}
       \item[6)] An arbitrary point $S$ $(x_s, y_s, z_s)$ of the secondary circular filament has parametric coordinates
   \begin{equation}\label{eq:coordinates S}
        \begin{array}{l}
          x_s=x_c+R_su_x\cos\vartheta+R_sv_x\sin\vartheta; \\
          y_s=y_c+R_su_y\cos\vartheta+R_sv_y\sin\vartheta;  \\
          z_s=z_c+R_su_z\cos\vartheta+R_sv_z\sin\vartheta, \;\vartheta\in[0,2\pi].
        \end{array}
       \end{equation}
       This is well-known parametric equation of circle in 3D space. The filamentary circular filaments are the part of this circle.
      \item[7)] The differential element of the secondary circular filament is given by,
       \begin{equation}\label{eq:differential coord of SC}
         d\vec{l}_s=R_s\{l_{xs}, l_{ys}, l_{zs}\}d\vartheta, \; \;\vartheta\in[0,2\pi],
       \end{equation}
       where
         \begin{equation}\label{eq:differential coord of SC comps}
         \begin{array}{l}
          l_{xs}= -u_x\sin\vartheta+v_x\cos\vartheta;\\
          l_{ys}= -u_y\sin\vartheta+v_y\cos\vartheta; \\
          l_{zs}=-u_z\sin\vartheta+v_z\cos\vartheta.
         \end{array}
       \end{equation}
 \end{description}

\subsection{Stiffness calculation}
To calculate the stiffness between two inclined circular loops as the prime interest of this article we use the analytical formulas for calculating the magnetic field produced by the primary current carrying with the current $I_p$ at the arbitrary point $S$ $(x_s, y_s, z_s)$ of the secondary inclined current carrying loop with the current $I_s$ \cite{Babic2021}. 
Hence, the components of the field can be calculated as follows,
\begin{equation}\label{eq:B-x}
B_{x}=\frac{\mu_{0} I_{p} z_{s} k}{16 \pi p^{2} \sqrt{R_{p} p}\left(1-k^{2}\right)} I_{x};
\end{equation}
\begin{equation}\label{eq:B-y}
B_{y}=\frac{\mu_{0} I_{p} z_{s} k}{16 \pi p^{2} \sqrt{R_{p} p}\left(1-k^{2}\right)} I_{y};
\end{equation}
\begin{equation}\label{eq:B-z}
B_{z}=-\frac{\mu_{0} I_{p} k}{16 \pi p \sqrt{R_{p} p}\left(1-k^{2}\right)} I_{z},
\end{equation}
where $p=\sqrt{x_s^2+y_s^2}$, $I_x=x_sA$, $I_y=y_sA$ and $I_z=D$ with
\begin{equation}\label{eq: A and D and k}
  \begin{array}{l}
     {\displaystyle k^2=\frac{4R_pp}{(R_p+p)^2+z_s^2}; }\\
    A=-2\left[(k^2-2)E(k)+(2-2k^2)K(k)\right]; \\
    D=-2\left[\left(k^2(R_p+p)-2p\right)E(k)+p\left(2-2k^2\right)K(k)\right].
  \end{array}\nonumber
\end{equation}
In given expressions $K(k)$  and $E(k)$ are the complete elliptic integrals of the first and the second kind, respectively.

Let us find the first derivatives of the components of the field with respect to coordinates $x_s$, $y_s$ and $z_s$, we can write
\begin{equation}\label{eq: first derivatives of B}
  \begin{array}{l}
     {\displaystyle \frac{\partial B_{x}}{\partial g}=\frac{\mu_{0} I_{p}}{16 \pi \sqrt{R_{p}}} \frac{p^{-\frac{3}{2}}}{\left(1-k^{2}\right)} \frac{z_{s}}{p}\left\{\left[-\frac{5}{2 p} k \frac{\partial p}{\partial g}+\frac{1+k^{2}}{1-k^{2}} \frac{\partial k}{\partial g}\right] I_{x}+k \frac{\partial I_{x}}{\partial g}\right\}; }\\
   {\displaystyle \frac{\partial B_{y}}{\partial g}=\frac{\mu_{0} I_{p}}{16 \pi \sqrt{R_{p}}} \frac{p^{-\frac{3}{2}}}{\left(1-k^{2}\right)} \frac{z_{s}}{p}\left\{\left[-\frac{5}{2 p} k \frac{\partial p}{\partial g}+\frac{1+k^{2}}{1-k^{2}} \frac{\partial k}{\partial g}\right] I_{y}+k \frac{\partial I_{y}}{\partial g}\right\}; }\\
    {\displaystyle \frac{\partial B_{z}}{\partial g}=\frac{\mu_{0} I_{p}}{16 \pi \sqrt{R_{p}}} \frac{p^{-\frac{3}{2}}}{\left(1-k^{2}\right)} \left\{\left[-\frac{3}{2 p} k \frac{\partial p}{\partial g}+\frac{1+k^{2}}{1-k^{2}} \frac{\partial k}{\partial g}\right] I_{z}+k \frac{\partial I_{z}}{\partial g}\right\};}\\
    g=x_s,y_s,z_s,
  \end{array}
\end{equation}
where
\begin{equation}\label{eq:first der of k and p }\nonumber
    \begin{array}{l}
        {\displaystyle \frac{\partial k}{\partial x_{s}}=\frac{x_{s} k^{3}}{8 R_{p} p^{3}}\left[R_{p}^{2}+z_{s}^{2}-p^{2}\right], \;\frac{\partial k}{\partial y_{s}}=\frac{y_{s} k^{3}}{8 R_{p} p^{3}}\left[R_{p}^{2}+z_{s}^{2}-p^{2}\right], \; \frac{\partial k}{\partial z_{s}}=-\frac{z_{s} k^{3}}{4 R_{p} p}; } \\
       {\displaystyle \frac{\partial p}{\partial x_{s}}=\frac{x_s}{p}, \;\frac{\partial p}{\partial y_{s}}=\frac{y_s}{p}, \; \frac{\partial p}{\partial z_{s}}=0; }
    \end{array}
\end{equation}
\begin{equation}\label{eq:first der of I }\nonumber
    \begin{array}{l}
        {\displaystyle \frac{\partial I_x}{\partial x_{s}}=A+Cx_s\frac{\partial k}{\partial x_{s}}, \;\frac{\partial I_x}{\partial y_{s}}=Cx_s\frac{\partial k}{\partial y_{s}}, \; \frac{\partial I_x}{\partial z_{s}}=Cx_s\frac{\partial k}{\partial z_{s}}; } \\
       {\displaystyle \frac{\partial I_y}{\partial x_{s}}=Cy_s\frac{\partial k}{\partial x_{s}}, \;\frac{\partial I_y}{\partial y_{s}}=A+Cy_s\frac{\partial k}{\partial y_{s}}, \; \frac{\partial I_y}{\partial z_{s}}=Cy_s\frac{\partial k}{\partial z_{s}}; }\\
       {\displaystyle \frac{\partial I_z}{\partial x_{s}}=D\frac{x_s}{p}=T, \;\frac{\partial I_z}{\partial y_{s}}=D\frac{y_s}{p}+T, \; \frac{\partial I_z}{\partial z_{s}}=T; }
    \end{array}
\end{equation}
\begin{equation}\label{eq:C and T}\nonumber
    \begin{array}{l}
        {\displaystyle C=-6x_sk\left[E(k)-K(k)\right]; } \\
       {\displaystyle T=-2k\frac{\partial k}{\partial x_s}\left[3(R_p+p)E(k)+(R_p+3p)K(k)\right]. }
    \end{array}
\end{equation}
Accounting for the fact that
\begin{equation}\label{eq:der of l}
  \frac{\partial l_{xs}}{\partial g}=\frac{\partial l_{ys}}{\partial g}=\frac{\partial l_{zs}}{\partial g}=0, \; g=x_s,y_s,z_s,
\end{equation}
magnetic stiffness calculation for the corresponding
force between two current-carrying circular filaments
arbitrarily oriented in the space is given by the following  formulas \cite{Babic2021}:
\begin{equation}\label{eq:Sxx}
 S_{x x}=-\frac{\partial F_{x}}{\partial x_{s}}=-I_{s} R_{s} \int_{0}^{2 \pi}\left[l_{y s} \frac{\partial B_{z}}{\partial x_{s}}-l_{z s} \frac{\partial B_{y}}{\partial x_{s}}\right] d \vartheta;
\end{equation}
\begin{equation}\label{eq:Sxy}
 S_{x y}=-\frac{\partial F_{x}}{\partial y_{s}}=-I_{s} R_{s} \int_{0}^{2 \pi}\left[l_{y s} \frac{\partial B_{z}}{\partial y_{s}}-l_{z s} \frac{\partial B_{y}}{\partial y_{s}}\right] d \vartheta;
\end{equation}
\begin{equation}\label{eq:Sxz}
 S_{x z}=-\frac{\partial F_{x}}{\partial z_{s}}=-I_{s} R_{s} \int_{0}^{2 \pi}\left[l_{y s} \frac{\partial B_{z}}{\partial z_{s}}-l_{z s} \frac{\partial B_{y}}{\partial z_{s}}\right] d \vartheta;
\end{equation}
\begin{equation}\label{eq:Syx}
 S_{y x}=-\frac{\partial F_{y}}{\partial x_{s}}=I_{s} R_{s} \int_{0}^{2 \pi}\left[l_{x s} \frac{\partial B_{z}}{\partial x_{s}}-l_{z s} \frac{\partial B_{x}}{\partial x_{s}}\right] d \vartheta;
\end{equation}
\begin{equation}\label{eq:Syy}
 S_{y y}=-\frac{\partial F_{y}}{\partial y_{s}}=I_{s} R_{s} \int_{0}^{2 \pi}\left[l_{x s} \frac{\partial B_{z}}{\partial y_{s}}-l_{z s} \frac{\partial B_{x}}{\partial y_{s}}\right] d \vartheta;
\end{equation}
\begin{equation}\label{eq:Syz}
 S_{y z}=-\frac{\partial F_{y}}{\partial z_{s}}=I_{s} R_{s} \int_{0}^{2 \pi}\left[l_{x s} \frac{\partial B_{z}}{\partial z_{s}}-l_{y s} \frac{\partial B_{x}}{\partial z_{s}}\right] d \vartheta;
\end{equation}
\begin{equation}\label{eq:Szx}
 S_{z x}=-\frac{\partial F_{z}}{\partial x_{s}}=-I_{s} R_{s} \int_{0}^{2 \pi}\left[l_{x s} \frac{\partial B_{y}}{\partial x_{s}}-l_{y s} \frac{\partial B_{x}}{\partial x_{s}}\right] d \vartheta;
\end{equation}
\begin{equation}\label{eq:Szy}
 S_{z y}=-\frac{\partial F_{z}}{\partial y_{s}}=-I_{s} R_{s} \int_{0}^{2 \pi}\left[l_{x s} \frac{\partial B_{y}}{\partial y_{s}}-l_{y s} \frac{\partial B_{x}}{\partial y_{s}}\right] d \vartheta;
\end{equation}
\begin{equation}\label{eq:Szz}
 S_{z z}=-\frac{\partial F_{z}}{\partial z_{s}}=-I_{s} R_{s} \int_{0}^{2 \pi}\left[l_{x s} \frac{\partial B_{y}}{\partial z_{s}}-l_{y s} \frac{\partial B_{x}}{\partial z_{s}}\right] d \vartheta.
\end{equation}

Thus, all magnetic stiffness components (\ref{eq:Sxx})-(\ref{eq:Szz}) between two inclined current-carrying loops are given in the simple integral form, over the complete elliptic integrals of the first and the second kind.
These expressions can be used for resolving the singular cases. It is necessary to use expressions (\ref{eq:Sxx})-(\ref{eq:Szz}) with the following conditions $(a=c=0,l=0,L= |b|)$ for the unit vectors $\vec{u}=\left\{-1, 0, 0\right\}$ and $\vec{v}=\left\{0, 0, 1\right\}$. The loops are perpendicular mutually.

%

It is clear that $S_{xy}=S_{yx}$, $S_{xz}=S_{zx}$, $S_{yz}=S_{zy}$, so that the calculation can be simplified by finding only sixth stiffness: $S_{xx}$, $S_{yy}$, $S_{zz}$,   $S_{xy}$, $S_{xz}$, and $S_{yz}$. Doing further investigation one can find that \cite{Paul2014},
\begin{equation}\label{eq:sum condition}
  S_{xx}+S_{yy}+ S_{zz}=0,
\end{equation}
so that the problem of the stiffness calculation can be limited to find only five components, for instance, $S_{xx}$, $S_{yy}$,  $S_{xy}$, $S_{xz}$, and $S_{yz}$.

\section{Mutual Inductance Method (MIM)}
In this section, the mutual inductance method as an alternative to Babic's method discussed above is presented. The essence of the method is that  the calculation of the stiffness of the corresponding  magnetic force  is performed by means of finding the second derivatives of the function of mutual inductance between two  circular filaments with respect to the appropriated coordinates.


\subsection{ Preliminary discussion}
 The general scheme of arbitrarily positioning of two current-carrying circular filaments with respect to each other is considered as shown in Fig. \ref{fig:scheme}. 
The linear misalignment of the secondary circle with respect to the primary one is defined by the coordinates of the centre $C$ ($x_c,y_c,z_c$).
The angular misalignment of the secondary circle can be defined by using Grover's  angles \cite[page 207]{Grover2004}. Namely, the angle of $\theta$ and $\eta$ corresponds to the angular rotation around an axis passing through the diameter of the secondary circle, and then the rotation of this axis lying on the surface $x'By'$ around the vertical $z'$ axis, respectively, as it is shown in Figure \ref{fig:angular position}(a). Accounting for Eq. (\ref{eq:unit vector N}),   these two angles have the following relationship with constants of inclined plane (\ref{eq:inclined plane}):
\begin{equation}\label{eq:relatioship bet angles and constants}
  \theta=\arccos\left(\frac{c}{L}\right),\;\eta=\arccos\left(\frac{-b}{L\sin\theta}\right).
\end{equation}
\begin{figure*}[!t]
    \centering
     \subfigure[]
    {
    \centering
        \includegraphics[width=1.8in]{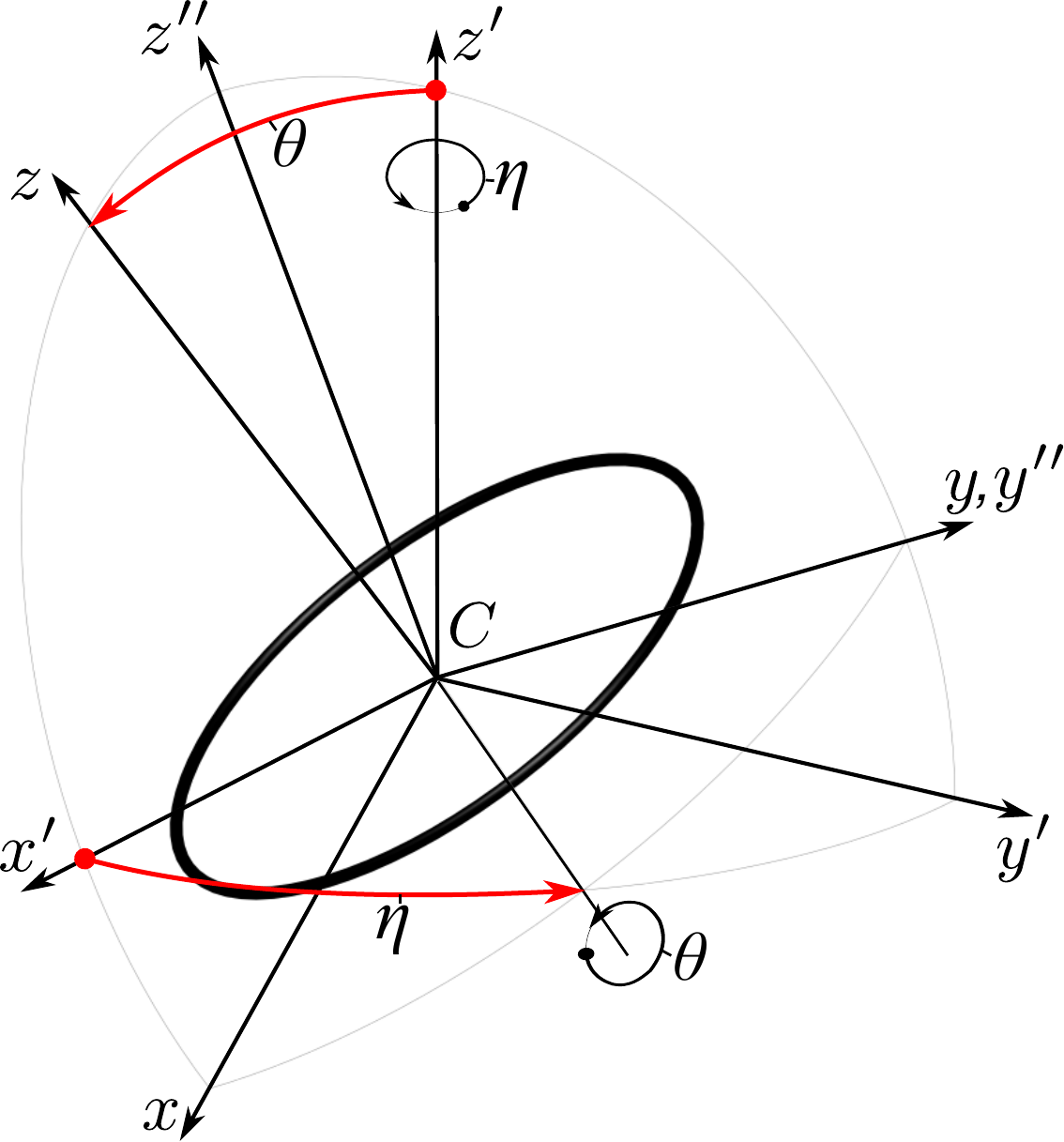}
       \label{fig:Grover angles}
        }\quad
        \subfigure[ ]
    {
    \centering
        \includegraphics[width=1.8in]{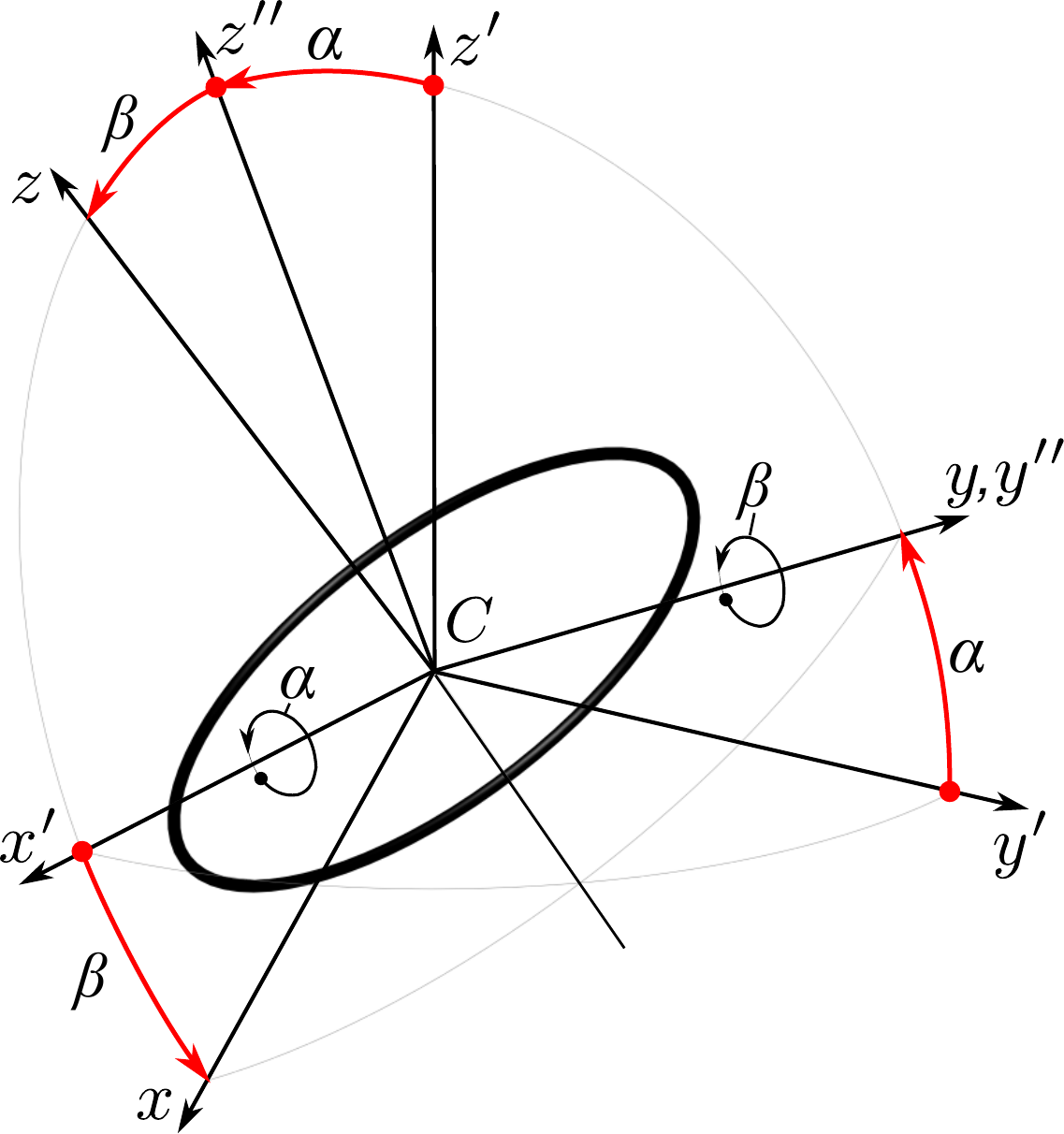}
       \label{fig:Bronyan}
        }\quad
    \caption{Two manners for determining the angular position of the secondary circle with respect to the primary one: $x'y'z'$ is the auxiliary CF the axes of which are parallel to the axes of $XYZ$, respectively; $x''y''z''$ is the auxiliary CF defined in such a way that the $x'$ and $x''$ are coincide, but the $z''$ and $y''$ axis is rotated by the $\alpha$ angle with respect to the $z'$ and $y'$ axis, respectively.  }
    \label{fig:angular position}
\end{figure*}

The same angular misalignment can be determined through the $\alpha$ and $\beta$ angle, which corresponds to the angular rotation around the $x'$ axis and then around the $y''$ axis, respectively, as it is shown in Figure \ref{fig:angular   position}(b). This additional second  manner  is more convenient in a case of study dynamics and stability issues, for instance, applying to axially symmetric inductive levitation systems \cite{Poletkin2014a,PoletkinLuWallrabeEtAl2017b} in compared with    Grover's manner.  These two pairs of angles have the following relationship with respect to each other such as \cite{Poletkin2019}:
\begin{equation}\label{eq:angles}
  \left\{\begin{array}{l}
   \sin\beta=\sin\eta\sin\theta;\\
   \cos\beta\sin\alpha=\cos\eta\sin\theta.
  \end{array}\right.
\end{equation}

Then, the mutual inductance between  two circular filaments can be calculated by
the following formulas, which were derived by using Kalantarov-Zeitlin's approach in work \cite{Poletkin2019} for two cases.
Introducing the following dimensionless coordinates:
  \begin{equation}\label{eq:dimensionless_par}
    {\displaystyle {x}=\frac{x_c}{R_s};\; {y}=\frac{y_c}{R_s};\; {z}=\frac{z_c}{R_s}; {s}=\sqrt{{x}^2+{y}^2},} \\
  \end{equation}
for the first case when the $\theta$ angle is lying in an interval of $0\leq\theta<\pi/2$, the formula can be written as
\begin{equation}\label{eq:NEW FORMULA}
  M=\frac{\mu_0\sqrt{R_pR_s}}{\pi}\int_{0}^{2\pi}{r}\cdot U\cdot\Phi(k)d\varphi,
\end{equation}
where
\begin{equation}\label{eq:r}
   {r}={r}(\theta,\eta)=\frac{\cos\theta}{\sqrt{\sin^2(\varphi-\eta)+\cos^2\theta\cos^2(\varphi-\eta)}},
\end{equation}
\begin{equation}\label{eq:U}
  U=U({x},{y},\theta,\eta)=\frac{R}{{\rho}^{1.5}}=\frac{{r}+t_1\cdot\cos\varphi+t_2\cdot\sin\varphi}{{\rho}^{1.5}},
\end{equation}
\begin{equation}\label{eq:t and rho}
  \begin{array}{l}
    t_1=t_1({x},{y},\theta,\eta)={x}+0.5{r}^2\tan^2\theta\sin(2(\varphi-\eta))\cdot{y}, \\
     t_2=t_2({x},{y},\theta,\eta)={y}-0.5{r}^2\tan^2\theta\sin(2(\varphi-\eta))\cdot{x},\\
    {\rho}={\rho}({x},{y},\theta,\eta) =\sqrt{{r}^2+2{r}\cdot \left({x}\cos(\varphi)+{y}\sin(\varphi)\right)+{s}^2},
  \end{array}
\end{equation}
\begin{equation}\label{eq:Phi}
   \Phi(k)=\frac{1}{k}\left[\left(1-\frac{k^2}{2}\right)K(k)-E(k)\right],
\end{equation}
and
\begin{equation}\label{eq:k}
\begin{array}{l}
   {\displaystyle k^2=k^2({x},{y},{z},\theta,\eta)=\frac{4\nu{\rho}}{(\nu{\rho}+1)^2+\nu^2{z}_{\lambda}^2},
}\\
 {\displaystyle \nu=R_s/R_p,\;  {z}_{\lambda}={z}+{r}\tan\theta\sin(\varphi-\eta)}.
\end{array}
\end{equation}
For  the second case when the $\theta$ angle is equal to $\pi/2$ and the two filament circles are mutually perpendicular to each other,  the formula becomes
\begin{equation}\label{eq:Singular case}
 \begin{array}{l}
  {\displaystyle M=\frac{\mu_0\sqrt{R_pR_s}}{\pi}\left\{\int_{-1}^{1}U\cdot\Phi(k)d\bar{\ell}\right.}\\
  {\displaystyle  \left.+\int_{1}^{-1}U\cdot\Phi(k)d\bar{\ell}\right\}},
  \end{array}
\end{equation}
where
\begin{equation}\label{eq:U sing c}
  U=U({x},{y},\eta)=\frac{R}{{\rho}^{1.5}}=\frac{t_1-t_2}{{\rho}^{1.5}},
\end{equation}
\begin{equation}\label{eq:t for singular case}
  \begin{array}{l}
    t_1=t_1({x},\eta)=\sin\eta\cdot({x}+\bar{\ell}\cos\eta), \\
     t_2=t_2({y},\eta)=\cos\eta\cdot({y}+\bar{\ell}\sin\eta),\\
     {\rho}={\rho}({x},{y},\eta) =\sqrt{{s}^2+2\bar{\ell}\cdot  \left({x}\cos(\eta)+{y}\sin(\eta)\right)+\bar{\ell}^2},
  \end{array}
\end{equation}
and $\bar{\ell}=\ell/R_s$ is the dimensionless variable. The functions $\Phi(k)$ and $k=k({x},{y},{z},\eta)$ in formula (\ref{eq:Singular case}) have the same structures as defined by Eq. (\ref{eq:Phi}) and (\ref{eq:k}), respectively. Besides that, in the elliptic module $k=k({x},{y},{z},\eta)$, the ${z}_{\lambda}$ function is governed as follows
\begin{equation}\label{eq: Z singular case}
  {z}_{\lambda}={z}\pm\sqrt{1-\bar{\ell}^2},
\end{equation}
Note that integrating formula (\ref{eq:Singular case}) between $-1$ and $1$, Eq. (\ref{eq: Z singular case}) is calculated with the positive sign and for the other direction the negative sign is taken.

\subsection{ Stiffness calculation}

Assuming that the primary and secondary circular filaments carry the currents of $I_p$ and $I_s$, respectively, hence, the magnetic stiffness corresponding to the force arising between these two current-carring circular filaments can be calculated by taking the second derivatives of the function of the magnetic energy stored in such the system with respect to the appropriate coordinates. Hence, all nine components of  the magnetic stiffness can be  calculated by
\begin{equation}\label{eq: general stiffness}
  {\displaystyle  S_{gq}=-I_pI_s\frac{\partial^2 M}{\partial g\partial q},}
\end{equation}
where $g,q=x_c$, $y_c$, or $z_c$.
 Thus, to derive formulas for calculation of the magnetic stiffness  between two arbitrarily oriented circular filaments, the second derivatives of formulas of mutual inductance, namely, represented by Eq. (\ref{eq:NEW FORMULA}) and (\ref{eq:Singular case})   must be taken. Similar to the calculation of magnetic force in such the filament system \cite{Poletkin2021a}, finding the second derivatives of mutual inductance  is reduced to taking the second
derivatives of their kernel functions.

  \subsubsection{For the case of $0\leq\theta<\pi/2$}  Formula (\ref{eq:NEW FORMULA}) for calculation of mutual inductance is considered. Its kernel is defined as
\begin{equation}\label{eq:kernel NF}
  \mathrm{Kr}={r}\cdot U\cdot\Phi(k).
\end{equation}

 According to the definitions of functions $r$, $U$, $\Phi(k)$ and $k$ given in Eq. (\ref{eq:r}), (\ref{eq:U}), (\ref{eq:Phi}) and (\ref{eq:k}), respectively, the second $x_c$-and $y_c$-derivative of kernel $ \mathrm{Kr}$ can be written as
\begin{equation}\label{eq:second g-der of kernel NF}
 \begin{array}{c}
   {\displaystyle  \frac{\partial^2\mathrm{Kr}}{\partial q^2}=\frac{\partial^2\mathrm{Kr}}{\partial g^2}\frac{1}{R_s^2}=\frac{r}{R_s^2}\cdot\left[\frac{\partial^2 U}{\partial g^2} \cdot\Phi(k)+2\frac{\partial U}{\partial g}\cdot\frac{d \Phi(k)}{d k} \cdot\frac{\partial k}{\partial g}\right.} \\
   {\displaystyle \left. +U\left(\frac{d^2 \Phi(k)}{d k^2} \cdot\left(\frac{\partial k}{\partial g}\right)^2+\frac{d \Phi(k)}{d k} \cdot\frac{\partial^2 k}{\partial g^2} \right)\right],}
 \end{array}
\end{equation}
where $q=x_c,y_c$, and  $g=x,y$,
\begin{equation}\label{eq:dU-g}
 \begin{array}{l}
  {\displaystyle  \frac{\partial U}{\partial g}=\left({\displaystyle\frac{\partial R}{\partial g}}\cdot\rho-1.5\cdot R\cdot\frac{\partial \rho}{\partial g}\right)\bigg/{{\rho}^{2.5}},}\\
  {\displaystyle  \frac{\partial^2 U}{\partial g^2}=\frac{{\displaystyle\left[-0.5\frac{\partial R}{\partial g}\frac{\partial \rho}{\partial g}-1.5R\frac{\partial^2 \rho}{\partial g^2}\right]\cdot\rho-2.5\cdot \left[{\displaystyle\frac{\partial R}{\partial g}}\rho-1.5 R\frac{\partial \rho}{\partial g}\right]\cdot\frac{\partial \rho}{\partial g}}}{{{\rho}^{3.5}}},}\\
  {\displaystyle \frac{\partial R}{\partial g}=\frac{\partial t_1}{\partial g}\cdot\cos\varphi+\frac{\partial t_2}{\partial g}\cdot\sin\varphi},\\
   \end{array}
\end{equation}
\begin{equation}\label{eq:dk-g}
  \begin{array}{l}
     {\displaystyle \frac{\partial k}{\partial g}=\frac{G}{H}\cdot\nu\frac{\partial \rho}{\partial g},
} \\
 {\displaystyle \frac{\partial^2 k}{\partial g^2}=\frac{ {\displaystyle\frac{\partial G}{\partial g}H-A\frac{\partial H}{\partial g}}}{H^2}\cdot\nu\frac{\partial \rho}{\partial g}+\frac{G}{H}\cdot\nu\frac{\partial^2 \rho}{\partial g^2},
}\\
  {\displaystyle G=2/k-k(\nu{\rho}+1),\;  H=(\nu{\rho}+1)^2+\nu^2{z}_{\lambda}^2,
}  \\
 {\displaystyle \frac{\partial G}{\partial g}=-\left[2/k^2+\nu{\rho}+1\right]\frac{\partial k}{\partial g}-k\cdot\nu\frac{\partial \rho}{\partial g}, \frac{\partial H}{\partial g}=2(\nu{\rho}+1)\nu\frac{\partial \rho}{\partial g},
}
  \end{array}
\end{equation}
\begin{equation}\label{eq:dPhi}
\begin{array}{l}
 {\displaystyle  \frac{d \Phi(k)}{d k}=\frac{1}{k^2}\left[\frac{2-k^2}{2(1-k^2)}E(k)-K(k)\right]},\\
 {\displaystyle \frac{d^2 \Phi(k)}{d k^2}=-\frac{\left(4-7k^2+k^4\right)E(k)+\left(-4+9k^2-5k^4\right)K(k)}{2k^3(k^2 -1)^2}},
\end{array}
\end{equation}
and  when $g=x$ we have
\begin{equation}\label{eq:g-x}
 \begin{array}{l}
   {\displaystyle \frac{\partial t_1}{\partial x}=1,\;\frac{\partial t_2}{\partial x}=-0.5{r}^2\tan^2\theta\sin(2(\varphi-\eta))},\\
    {\displaystyle \frac{\partial \rho}{\partial x}=\left(r\cdot\cos\varphi +x\right)\big/{{\rho}},\; \frac{\partial^2 \rho}{\partial x^2}=\left(\rho-(r\cdot\cos\varphi +x)\frac{\partial \rho}{\partial x}\right)\big/{{\rho}^2},}
   \end{array}
\end{equation}
and  when $g=y$ we have
\begin{equation}\label{eq:g-y}
 \begin{array}{l}
   {\displaystyle \frac{\partial t_1}{\partial y}= 0.5{r}^2\tan^2\theta\sin(2(\varphi-\eta)),\;\frac{\partial t_2}{\partial y}=1},\\
    {\displaystyle \frac{\partial \rho}{\partial y}=\left(r\cdot\sin\varphi +y\right)\big/{{\rho}}, \; \frac{\partial^2 \rho}{\partial y^2}=\left(\rho-(r\cdot\sin\varphi +y)\frac{\partial \rho}{\partial y}\right)\big/{{\rho}^2}.}
   \end{array}
\end{equation}
The second $z_c$-derivative of kernel $ \mathrm{Kr}$ is
\begin{equation}\label{eq:second z-der of kernel NF}
   {\displaystyle  \frac{\partial^2\mathrm{Kr}}{\partial z_c^2}=\frac{\partial^2\mathrm{Kr}}{\partial z^2}\frac{1}{R_s^2}=\frac{r}{R_s^2}\cdot U\left[\frac{d^2 \Phi(k)}{d k^2} \cdot\left(\frac{\partial k}{\partial z}\right)^2+\frac{d \Phi(k)}{d k} \cdot\frac{\partial^2 k}{\partial z^2} \right],}
\end{equation}
where
\begin{equation}\label{eq:dk-z}
\begin{array}{l}
   {\displaystyle \frac{\partial k}{\partial z}=-\sqrt{4\nu\rho}\cdot\frac{\nu^2{z}_{\lambda}}{\left((\nu{\rho}+1)^2+\nu^2{z}_{\lambda}^2\right)^{3/2}}
},\\
{\displaystyle \frac{\partial^2 k}{\partial z^2}=\nu^2\sqrt{4\nu\rho}\cdot\frac{2\nu^2{z}_{\lambda}^2-(\nu{\rho}+1)^2}{\left((\nu{\rho}+1)^2+\nu^2{z}_{\lambda}^2\right)^{5/2}}
}.
 \end{array}
\end{equation}
Note that in Eq. (\ref{eq:second z-der of kernel NF}), the first and second derivative of function $\Phi$ with respect to $k$ are the same as in Eqs (\ref{eq:dPhi}), respectively.

For further differentiation of the kernel, it is recognized that
\begin{equation}\label{eq:second gz- and xyder of kernel NF}
 {\displaystyle  \frac{\partial^2\mathrm{Kr}}{\partial q\partial z_c}=\frac{\partial^2\mathrm{Kr}}{\partial z_c\partial q }, \; \frac{\partial^2\mathrm{Kr}}{\partial x_c\partial y_c}=\frac{\partial^2\mathrm{Kr}}{\partial y_c\partial x_c },}
\end{equation}
where $q=x_c,y_c$. Using  properties (\ref{eq:second gz- and xyder of kernel NF}) the derivation of the second derivatives, where the variable $z_c$ is involved, can be simplified by taking the first derivative of the kernel with respect to $z_c$.   Following this, we have
\begin{equation}\label{eq:second gz-der of kernel NF}
\begin{array}{c}
  {\displaystyle  \frac{\partial^2\mathrm{Kr}}{\partial q\partial z_c}=\frac{\partial^2\mathrm{Kr}}{\partial g\partial z}\frac{1}{R_s^2}=\frac{r}{R_s^2}\cdot\left[\frac{\partial U}{\partial g}\cdot\frac{d \Phi(k)}{d k} \cdot\frac{\partial k}{\partial z} \right.} \\
   {\displaystyle  \left. +U  \left(\frac{d^2 \Phi(k)}{d k^2} \cdot\frac{\partial k}{\partial g}\cdot\frac{\partial k}{\partial z}+\frac{d \Phi(k)}{d k} \cdot\frac{\partial^2 k}{\partial g\partial z}\right)\right],          }
\end{array}
\end{equation}
where $q=x_c,y_c$ and $g=x,y$,
\begin{equation}\label{eq:dk-gz}
 \begin{array}{l}
{\displaystyle \frac{\partial k}{\partial z}=-{2}\cdot\nu^{2.5}{z}_{\lambda}\cdot\frac{\sqrt{\rho}}{H^{3/2}}
},\\
 {\displaystyle \frac{\partial^2 k}{\partial g \partial z}= \frac{-\nu^{2.5}{z}_{\lambda}}{\sqrt{\rho}}\cdot\frac{{\displaystyle \frac{\partial \rho}{\partial g}\cdot H-3{\rho}\cdot\frac{\partial H}{\partial g}}}{H^{5/2}},
}\\
  {\displaystyle  H=(\nu{\rho}+1)^2+\nu^2{z}_{\lambda}^2,\;\frac{\partial H}{\partial g}=2(\nu{\rho}+1)\nu\frac{\partial \rho}{\partial g}.
}
  \end{array}
\end{equation}
The second derivative with respect to variables $x_c$ and $y_c$ is
\begin{equation}\label{eq:second xy-der of kernel NF}
\begin{array}{c}
  {\displaystyle  \frac{\partial^2\mathrm{Kr}}{\partial x_c\partial y_c}=\frac{\partial^2\mathrm{Kr}}{\partial x\partial y}\frac{1}{R_s^2}=\frac{r}{R_s^2}\cdot\left[\frac{\partial^2 U}{\partial x \partial y}\Phi(k)+\frac{\partial U}{\partial x}\cdot\frac{d \Phi(k)}{d k} \cdot\frac{\partial k}{\partial y} \right.} \\
   {\displaystyle  \left. +\frac{\partial U}{\partial y}\cdot\frac{d \Phi(k)}{d k} \cdot\frac{\partial k}{\partial x}+U  \left(\frac{d^2 \Phi(k)}{d k^2} \cdot\frac{\partial k}{\partial x}\cdot\frac{\partial k}{\partial y}+\frac{d \Phi(k)}{d k} \cdot\frac{\partial^2 k}{\partial x\partial y}\right)\right],          }
\end{array}
\end{equation}
where
\begin{equation}\label{eq:dU-xy}
 \begin{array}{l}
    {\displaystyle  \frac{\partial^2 U}{\partial y \partial x}=\left(\frac{\partial G_x}{\partial y}\rho-2.5G_x\frac{\partial \rho}{\partial y}\right)\bigg/{{\rho}^{3.5}},}\\
    {\displaystyle    G_x={\displaystyle\frac{\partial R}{\partial x}}\cdot\rho-1.5\cdot R\cdot\frac{\partial \rho}{\partial x},}\\
  {\displaystyle \frac{\partial G_x}{\partial y}={\displaystyle\frac{\partial R}{\partial x}}\cdot\frac{\partial \rho}{\partial y}-1.5\cdot {\displaystyle\frac{\partial R}{\partial y}}\cdot\frac{\partial \rho}{\partial x}-1.5R\frac{\partial^2 \rho}{\partial y \partial x},}\\
  {\displaystyle\frac{\partial^2 \rho}{\partial y \partial x}=-\frac{\partial \rho}{\partial x}\cdot\frac{\partial \rho}{\partial y}\cdot\frac{1}{\rho}},
   \end{array}
\end{equation}
\begin{equation}\label{eq:dk-xy}
  \begin{array}{l}
 {\displaystyle \frac{\partial^2 k}{\partial y \partial x}=\frac{ {\displaystyle\frac{\partial G}{\partial y}H-G\frac{\partial H}{\partial y}}}{H^2}\cdot\nu\frac{\partial \rho}{\partial x}+\frac{G}{H}\cdot\nu\frac{\partial^2 \rho}{\partial y \partial x},
}\\
  {\displaystyle G=2/k-k(\nu{\rho}+1),\;  H=(\nu{\rho}+1)^2+\nu^2{z}_{\lambda}^2,
}  \\
 {\displaystyle \frac{\partial G}{\partial y}=-\left[2/k^2+\nu{\rho}+1\right]\frac{\partial k}{\partial y}-k\cdot\nu\frac{\partial \rho}{\partial y}, \frac{\partial H}{\partial y}=2(\nu{\rho}+1)\nu\frac{\partial \rho}{\partial y}.
}
  \end{array}
\end{equation}
Other derivatives of functions $\Phi$, $k$, $U$ in Eqs  (\ref{eq:second gz-der of kernel NF}) and (\ref{eq:second xy-der of kernel NF}) are defined in the same way as it has been shown above.

Hence, for this particular case when $0\leq\theta<\pi/2$, according to Eqs (\ref{eq:second g-der of kernel NF}), (\ref{eq:second z-der of kernel NF}), (\ref{eq:second gz-der of kernel NF}) and (\ref{eq:second xy-der of kernel NF}) all nine components of magnetic stiffness can be calculated as follows:
\begin{equation}\label{eq:Sqq NC}
 \begin{array}{l}
   {\displaystyle S_{qq}=-\frac{\mu_0 I_pI_s\sqrt{R_p}}{\pi R_s^{3/2}}\int_{0}^{2\pi}{r}\cdot \left[\frac{\partial^2 U}{\partial g^2} \cdot\Phi(k)+2\frac{\partial U}{\partial g}\cdot\frac{d \Phi(k)}{d k} \cdot\frac{\partial k}{\partial g}\right.} \\
   {\displaystyle \left. +U\left(\frac{d^2 \Phi(k)}{d k^2} \cdot\left(\frac{\partial k}{\partial g}\right)^2+\frac{d \Phi(k)}{d k} \cdot\frac{\partial^2 k}{\partial g^2} \right)\right]d\varphi;}
 \end{array}
\end{equation}
\begin{equation}\label{eq:Szz NC}
 \begin{array}{l}
   {\displaystyle S_{z_cz_c}=-\frac{\mu_0 I_pI_s\sqrt{R_p}}{\pi R_s^{3/2}}\int_{0}^{2\pi}{r}\cdot U\left[\frac{d^2 \Phi(k)}{d k^2} \cdot\left(\frac{\partial k}{\partial z}\right)^2+\frac{d \Phi(k)}{d k} \cdot\frac{\partial^2 k}{\partial z^2} \right]d\varphi;}
 \end{array}
\end{equation}
\begin{equation}\label{eq:Sqz NC}
 \begin{array}{l}
   {\displaystyle S_{z_cq}= S_{qz_c}=-\frac{\mu_0 I_pI_s\sqrt{R_p}}{\pi R_s^{3/2}}\int_{0}^{2\pi}{r}\cdot\left[\frac{\partial U}{\partial g}\cdot\frac{d \Phi(k)}{d k} \cdot\frac{\partial k}{\partial z} \right.} \\
   {\displaystyle \left. +U  \left(\frac{d^2 \Phi(k)}{d k^2} \cdot\frac{\partial k}{\partial g}\cdot\frac{\partial k}{\partial z}+\frac{d \Phi(k)}{d k} \cdot\frac{\partial^2 k}{\partial g\partial z}\right)\right]d\varphi;}
 \end{array}
\end{equation}
\begin{equation}\label{eq:Sxy NC}
 \begin{array}{l}
   {\displaystyle S_{x_cy_c}= S_{y_cx_c}=-\frac{\mu_0 I_pI_s\sqrt{R_p}}{\pi R_s^{3/2}}\int_{0}^{2\pi}r\cdot\left[\frac{\partial^2 U}{\partial x \partial y}\Phi(k)+\frac{\partial U}{\partial x}\cdot\frac{d \Phi(k)}{d k} \cdot\frac{\partial k}{\partial y} \right.} \\
   {\displaystyle  \left. +\frac{\partial U}{\partial y}\cdot\frac{d \Phi(k)}{d k} \cdot\frac{\partial k}{\partial x}+U  \left(\frac{d^2 \Phi(k)}{d k^2} \cdot\frac{\partial k}{\partial x}\cdot\frac{\partial k}{\partial y}+\frac{d \Phi(k)}{d k} \cdot\frac{\partial^2 k}{\partial x\partial y}\right)\right]d\varphi,}
 \end{array}
\end{equation}
where  $q=x_c$ or $y_c$, $g=x$ or $y$, respectively.

\subsubsection{The second case: $\theta=\pi/2$}
For this case, formula (\ref{eq:Singular case}) for calculation of mutual inductance is used. Its kernel is defined as
\begin{equation}\label{eq:kernel SC}
  \mathrm{Kr}=U\cdot\Phi(k).
\end{equation}
Accounting for that in this case the function $U$ is defined by Eq. (\ref{eq:U sing c}), then the second $x_c$-and $y_c$-derivative of kernel $ \mathrm{Kr}$ can be written as
\begin{equation}\label{eq:second g-der of kernel SC}
 \begin{array}{c}
   {\displaystyle  \frac{\partial^2\mathrm{Kr}}{\partial q^2}=\frac{\partial^2\mathrm{Kr}}{\partial g^2}\frac{1}{R_s^2}=\frac{1}{R_s^2}\cdot\left[\frac{\partial^2 U}{\partial g^2} \cdot\Phi(k)+2\frac{\partial U}{\partial g}\cdot\frac{d \Phi(k)}{d k} \cdot\frac{\partial k}{\partial g}\right.} \\
   {\displaystyle \left. +U\left(\frac{d^2 \Phi(k)}{d k^2} \cdot\left(\frac{\partial k}{\partial g}\right)^2+\frac{d \Phi(k)}{d k} \cdot\frac{\partial^2 k}{\partial g^2} \right)\right],}
 \end{array}
\end{equation}
where $q=x_c,y_c$, and  $g=x,y$. The first and second  derivatives of  function $U$ with respect to $g$ are taken analogically  as  shown in Eqs (\ref{eq:dU-g}). Similar to Eqs (\ref{eq:dk-g}), the derivatives of function $k$ are defined.
When $g=x$ we have
\begin{equation}\label{eq:g-x SC}
 \begin{array}{l}
   {\displaystyle \frac{\partial R}{\partial x}=\frac{\partial t_1}{\partial x}=\sin\eta, \frac{\partial \rho}{\partial x}=\frac{x+\bar{\ell}\cos\eta}{\rho}}\\
   {\displaystyle \frac{\partial^2 \rho}{\partial x^2}=\frac{\rho-(x+\bar{\ell}\cos\eta){\displaystyle\frac{\partial \rho}{\partial x}}}{\rho^2}},
   \end{array}
\end{equation}
and  when $g=y$ we can write
\begin{equation}\label{eq:g-y SC}
\begin{array}{l}
   {\displaystyle \frac{\partial R}{\partial y}=\frac{\partial t_2}{\partial y}=-\cos\eta, \frac{\partial \rho}{\partial y}=\frac{y+\bar{\ell}\sin\eta}{\rho}}\\
   {\displaystyle \frac{\partial^2 \rho}{\partial y^2}=\frac{\rho-(y+\bar{\ell}\sin\eta){\displaystyle\frac{\partial \rho}{\partial y}}}{\rho^2}}.
   \end{array}
\end{equation}
The second $z_c$-derivative of the kernel $ \mathrm{Kr}$ is defined similar to Eq. (\ref{eq:second z-der of kernel NF}) as 
\begin{equation}\label{eq:second z-der of kernel ЫС}
   {\displaystyle  \frac{\partial^2\mathrm{Kr}}{\partial z_c^2}=\frac{\partial^2\mathrm{Kr}}{\partial z^2}\frac{1}{R_s^2}=\frac{1}{R_s^2}\cdot U\left[\frac{d^2 \Phi(k)}{d k^2} \cdot\left(\frac{\partial k}{\partial z}\right)^2+\frac{d \Phi(k)}{d k} \cdot\frac{\partial^2 k}{\partial z^2} \right],}
\end{equation}
where the derivatives of function $k$ have the same meaning as in Eqs. (\ref{eq:dk-z}). Using  properties (\ref{eq:second gz- and xyder of kernel NF}) the derivation of the second derivatives with respect to the variables $z_c$ and $x_c$,  and  $z_c$ and $y_c$  can be written similar to Eq. (\ref{eq:second gz-der of kernel NF}) as follows
\begin{equation}\label{eq:second gz-der of kernel SC}
\begin{array}{c}
  {\displaystyle  \frac{\partial^2\mathrm{Kr}}{\partial q\partial z_c}=\frac{\partial^2\mathrm{Kr}}{\partial g\partial z}\frac{1}{R_s^2}=\frac{1}{R_s^2}\cdot\left[\frac{\partial U}{\partial g}\cdot\frac{d \Phi(k)}{d k} \cdot\frac{\partial k}{\partial z} \right.} \\
   {\displaystyle  \left. +U  \left(\frac{d^2 \Phi(k)}{d k^2} \cdot\frac{\partial k}{\partial g}\cdot\frac{\partial k}{\partial z}+\frac{d \Phi(k)}{d k} \cdot\frac{\partial^2 k}{\partial g\partial z}\right)\right],          }
\end{array}
\end{equation}
where   $q=x_c,y_c$ and $g=x,y$, taking into account Eqs. (\ref{eq:g-x SC}) and (\ref{eq:g-y SC}) the first and the second derivatives of $U$, $k$ are determined by Eqs. (\ref{eq:dU-g}) and (\ref{eq:dk-gz}), respectively.

The second derivative of the kernel with respect to variables $x_c$ and $y_c$ is defined in a similar way to Eq. (\ref{eq:second xy-der of kernel NF}) and is as follows
\begin{equation}\label{eq:second xy-der of kernel SC}
\begin{array}{c}
  {\displaystyle  \frac{\partial^2\mathrm{Kr}}{\partial x_c\partial y_c}=\frac{\partial^2\mathrm{Kr}}{\partial x\partial y}\frac{1}{R_s^2}=\frac{1}{R_s^2}\cdot\left[\frac{\partial^2 U}{\partial x \partial y}\Phi(k)+\frac{\partial U}{\partial x}\cdot\frac{d \Phi(k)}{d k} \cdot\frac{\partial k}{\partial y} \right.} \\
   {\displaystyle  \left. +\frac{\partial U}{\partial y}\cdot\frac{d \Phi(k)}{d k} \cdot\frac{\partial k}{\partial x}+U  \left(\frac{d^2 \Phi(k)}{d k^2} \cdot\frac{\partial k}{\partial x}\cdot\frac{\partial k}{\partial y}+\frac{d \Phi(k)}{d k} \cdot\frac{\partial^2 k}{\partial x\partial y}\right)\right].          }
\end{array}
\end{equation}
In Eq. (\ref{eq:second xy-der of kernel SC}) all derivatives of functions $U$ and $k$ are determined in Eq. (\ref{eq:second xy-der of kernel NF}), but the definitions of function $R$ and $\rho$  and their respective derivatives given in Eqs. (\ref{eq:g-x SC}) and (\ref{eq:g-y SC}) must be taken into account.

Hence, for this case when $\theta=\pi/2$, according to Eqs (\ref{eq:second g-der of kernel SC}),  (\ref{eq:second gz-der of kernel SC}) and (\ref{eq:second xy-der of kernel SC}) all nine components of magnetic stiffness can be calculated as follows:
\begin{equation}\label{eq:Sqq SC}
 \begin{array}{l}
   {\displaystyle S_{qq}=-\frac{\mu_0 I_pI_s\sqrt{R_p}}{\pi R_s^{3/2}}\left[\int_{-1}^{1}\frac{\partial^2 U}{\partial g^2} \cdot\Phi(k)+2\frac{\partial U}{\partial g}\cdot\frac{d \Phi(k)}{d k} \cdot\frac{\partial k}{\partial g}\right.} \\
   {\displaystyle  +U\left(\frac{d^2 \Phi(k)}{d k^2} \cdot\left(\frac{\partial k}{\partial g}\right)^2+\frac{d \Phi(k)}{d k} \cdot\frac{\partial^2 k}{\partial g^2} \right)d\bar{\ell}+\int_{1}^{-1}\frac{\partial^2 U}{\partial g^2} \cdot\Phi(k) }\\
  {\displaystyle \left. +2\frac{\partial U}{\partial g}\cdot\frac{d \Phi(k)}{d k} \cdot\frac{\partial k}{\partial g}+U\left(\frac{d^2 \Phi(k)}{d k^2} \cdot\left(\frac{\partial k}{\partial g}\right)^2+\frac{d \Phi(k)}{d k} \cdot\frac{\partial^2 k}{\partial g^2} \right)d\bar{\ell}\right]};
 \end{array}
\end{equation}
\begin{equation}\label{eq:Szz SC}
 \begin{array}{l}
   {\displaystyle S_{z_cz_c}=-\frac{\mu_0 I_pI_s\sqrt{R_p}}{\pi R_s^{3/2}}\left\{\int_{-1}^{1} U\left[\frac{d^2 \Phi(k)}{d k^2} \cdot\left(\frac{\partial k}{\partial z}\right)^2+\frac{d \Phi(k)}{d k} \cdot\frac{\partial^2 k}{\partial z^2} \right]d\bar{\ell}\right.}\\
  {\displaystyle \left. +\int_{1}^{-1} U\left[\frac{d^2 \Phi(k)}{d k^2} \cdot\left(\frac{\partial k}{\partial z}\right)^2+\frac{d \Phi(k)}{d k} \cdot\frac{\partial^2 k}{\partial z^2} \right]d\bar{\ell}\right\}};
 \end{array}
\end{equation}
\begin{equation}\label{eq:Sqz SC}
 \begin{array}{l}
   {\displaystyle S_{z_cq}= S_{qz_c}=-\frac{\mu_0 I_pI_s\sqrt{R_p}}{\pi R_s^{3/2}}\left[\int_{-1}^{1}\frac{\partial U}{\partial g}\cdot\frac{d \Phi(k)}{d k} \cdot\frac{\partial k}{\partial z} \right.} \\
   {\displaystyle  +U  \left(\frac{d^2 \Phi(k)}{d k^2} \cdot\frac{\partial k}{\partial g}\cdot\frac{\partial k}{\partial z}+\frac{d \Phi(k)}{d k} \cdot\frac{\partial^2 k}{\partial g\partial z}\right)d\bar{\ell}+\int_{1}^{-1}\frac{\partial U}{\partial g}\cdot\frac{d \Phi(k)}{d k} \cdot\frac{\partial k}{\partial z}}\\
     {\displaystyle \left. +U  \left(\frac{d^2 \Phi(k)}{d k^2} \cdot\frac{\partial k}{\partial g}\cdot\frac{\partial k}{\partial z}+\frac{d \Phi(k)}{d k} \cdot\frac{\partial^2 k}{\partial g\partial z}\right)d\bar{\ell}\right];}
 \end{array}
\end{equation}
\begin{equation}\label{eq:Sxy SC}
 \begin{array}{l}
   {\displaystyle S_{x_cy_c}= S_{y_cx_c}=-\frac{\mu_0 I_pI_s\sqrt{R_p}}{\pi R_s^{3/2}}\left[\int_{-1}^{1}\frac{\partial^2 U}{\partial x \partial y}\Phi(k)+\frac{\partial U}{\partial x}\cdot\frac{d \Phi(k)}{d k} \cdot\frac{\partial k}{\partial y} \right.} \\
   {\displaystyle  +\frac{\partial U}{\partial y}\cdot\frac{d \Phi(k)}{d k} \cdot\frac{\partial k}{\partial x}+U  \left(\frac{d^2 \Phi(k)}{d k^2} \cdot\frac{\partial k}{\partial x}\cdot\frac{\partial k}{\partial y}+\frac{d \Phi(k)}{d k} \cdot\frac{\partial^2 k}{\partial x\partial y}\right)d\bar{\ell}}\\
    {\displaystyle +\int_{1}^{-1}\frac{\partial^2 U}{\partial x \partial y}\Phi(k)+\frac{\partial U}{\partial x}\cdot\frac{d \Phi(k)}{d k} \cdot\frac{\partial k}{\partial y}+\frac{\partial U}{\partial y}\cdot\frac{d \Phi(k)}{d k} \cdot\frac{\partial k}{\partial x} } \\
    {\displaystyle  \left.+U  \left(\frac{d^2 \Phi(k)}{d k^2} \cdot\frac{\partial k}{\partial x}\cdot\frac{\partial k}{\partial y}+\frac{d \Phi(k)}{d k} \cdot\frac{\partial^2 k}{\partial x\partial y}\right)d\bar{\ell}\right],}
 \end{array}
\end{equation}
where  $q=x_c$ or $y_c$, $g=x$ or $y$, respectively.

Thus, the set of formulas (\ref{eq:Sqq NC})-(\ref{eq:Sxy NC}) and (\ref{eq:Sqq SC})-(\ref{eq:Sxy SC})  for calculation of all nine components  of magnetic stiffness of the corresponding
force arising between two current-carrying circular filaments
arbitrarily oriented in the space are derived by using the mutual inductance method. The derived formulas are intuitively understandable for application, they can be easily programmed.  For this purpose, the \textit{Matlab} language was used. The \textit{Matlab} files with the implemented formulas  are available   as   supplementary materials to this article.

\section{Numerical verification of derived formulas}
Developed  sets of formulas for calculation of nine components of magnetic stiffness of corresponding force between two current-carrying circular filaments derived  by means of  Babic's method Eqs (\ref{eq:Sxx})-(\ref{eq:Szz}) and the method of mutual inductance (Kalantarov-Zeitlin's method) Eqs (\ref{eq:Sqq NC})-(\ref{eq:Sxy NC}) and Eqs (\ref{eq:Sqq SC})-(\ref{eq:Sxy SC})  are mutually verified to each other through applying developed formulas to a number of examples designed in this section.
In all examples bellow, it is assumed that the carrying currents in both coils are equal to one ampere ($I_p=I_s=\SI{1}{\ampere}$). In addition to the calculation of components of magnetic stiffness of the considered filament system is supported by the set of expressions (\ref{eq:Srho GF})-(\ref{eq:Sdrho GF})  derived from   Grover's  formula for calculation of mutual inductance  \cite[page 207, Eq. (179)]{Grover2004}. The derivations of these expressions are shown in \ref{app:Grover}.
All  calculations for considered cases proved the robustness and efficiency of developed formulas.

\begin{figure}[!t]
  \centering
  \includegraphics[width=2.8in]{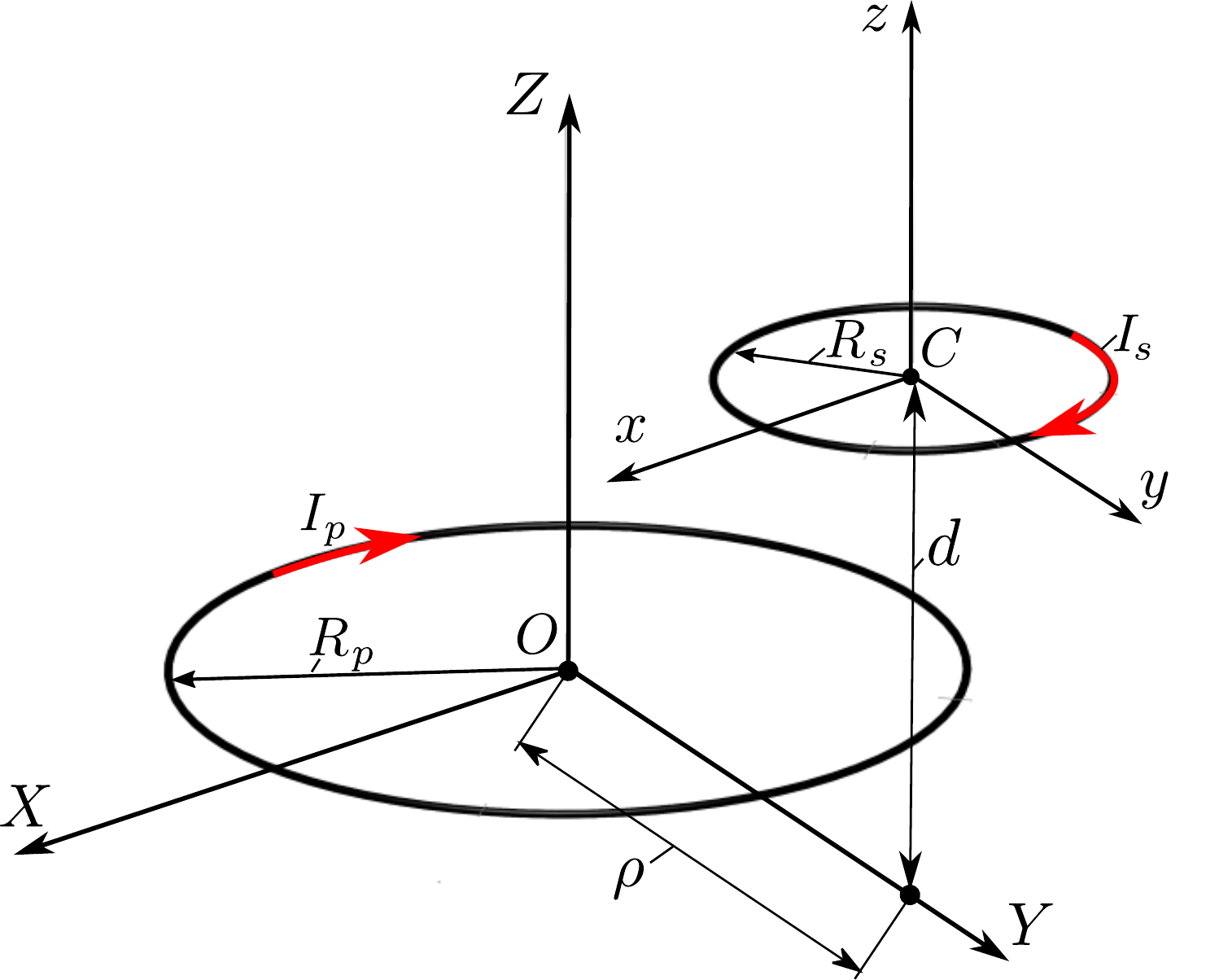}
  \caption{Geometrical scheme  of  circular  filaments with parallel axes denoted via Grover's notation: $\rho$ is the distance between axes; $d$ is the distance between the coils' planes $d=z_c$.   }\label{fig:filaments parallel axes}
\end{figure}
\subsection{Magnetic stiffness between circular  filaments with parallel axes}
The scheme for calculation of magnetic stiffness between circular  filaments with parallel
axes is shown in Fig. \ref{fig:filaments parallel axes}. The linear misalignment
in the Grover notation can be defined by the geometrical parameter, $d$, which is the distance between the planes of
circles  and the parameter, $\rho$,
is the distance between their axes.
These parameters have the following relationship to the notation defined in this article, namely, $z_c=d$ and $\rho=\sqrt{x_c^2+y_c^2}$. Fig. \ref{fig:filaments parallel axes} shows the particular case, when $\rho=y_c$. 

\subsubsection*{ Example 1: (Example 16, page 74 in Babi\v{c}'s work \cite{Babic2021})}\label{sec:exmple1}
Two coaxial circular filaments for which the primary coil has a radius of $R_p = 2$ \si{\metre}, and the secondary one $R_s = 1$ \si{\metre}. The axial distance between filaments is $d=z_c=1$\si{\metre}. The results of
calculation of diagonal and non-diagonal components of magnetic stiffness are as follows

\vspace*{1.0em}
\begin{footnotesize}
\begin{tabular}{lccc}
 \toprule
 &  BM, \si{\newton\per\metre}   &MIM, \si{\newton\per\metre}& GM, \si{\newton\per\metre} \\
   &  Eqs (\ref{eq:Sxx}), (\ref{eq:Syy}), (\ref{eq:Szz}) & Eqs (\ref{eq:Sqq NC}), (\ref{eq:Szz NC}) & Eqs (\ref{eq:Srho GF}), (\ref{eq:Sd GF}) \\
    \midrule
  $S_{xx}$ &   $1.032010586220236\times10^{-7} $ & $1.032010586220216\times10^{-7} $& $1.032010586220211\times10^{-7} $  \\
   $S_{yy}$ &  $1.032010586220238\times10^{-7} $& $1.032010586220216\times10^{-7} $ &  $1.032010586220211\times10^{-7} $  \\
    $S_{zz}$ &  $-2.064021172440475\times10^{-7} $  & $-2.064021172440485\times10^{-7} $ & $-2.064021172440499\times10^{-7}$ \\
  \toprule
\end{tabular}
\end{footnotesize}

\vspace*{1.0em}
\begin{footnotesize}
\begin{tabular}{lccc}
 \toprule
 &  BM, \si{\newton\per\metre}   &MIM, \si{\newton\per\metre}& GM, \si{\newton\per\metre} \\
   &  Eqs (\ref{eq:Sxy}), (\ref{eq:Syz}), (\ref{eq:Sxz}) & Eqs (\ref{eq:Sqz NC}), (\ref{eq:Sxy NC}) & Eqs (\ref{eq:Sdrho GF})\\
    \midrule
  $S_{xy}$ &  $-1.346233158563486\times10^{-142} $ & $8.298183971090421\times10^{-23} $&Not Applicable (NA) \\
   $S_{yz}$ &  $2.184515917637124\times10^{-142} $& $-6.010510193398827\times10^{-23} $ &  $1.739957438966064\times10^{-23} $  \\
    $S_{xz}$ &  $-2.533487747768601\times10^{-142} $  & $-5.887846720064157\times10^{-23} $ & $1.739957438966064\times10^{-23}$ \\
  \toprule
\end{tabular}
\end{footnotesize}

Analysis of the results of calculation shows that they are agree well to each other. The difference  arises after thirteenth digit in a resulting number. In the calculation of non-diagonal components, the  order of magnitude corresponding to minus twenty three can be considered as approximately equal zero. In Babic's method alternatively to  Eqs (\ref{eq:Sxy}), (\ref{eq:Syz}), (\ref{eq:Sxz}) the other set of formulas, namely, Eqs (\ref{eq:Syx}), (\ref{eq:Szy}), (\ref{eq:Szx}) can be used for calculation of non-diagonal components. The results are the same and equal to zeros. Also, note that the sum of diagonal components in all methods is almost zero and the condition (\ref{eq:sum condition}) is held with an accuracy of minus twenty one   order of magnitude. Worth noting that in some cases this condition helps us also to restore the missed component of stiffness in the orthogonal direction in Grover's method.

\subsubsection*{ Example 2}\label{sec:exmple2}
Let us consider the coils having the same radii as in the previous example 1, but the center of the secondary coil is located at the point $x_c=0$\si{\metre}, $y_c=0.5$\si{\metre} and $z_c=1$\si{\metre}, which corresponds to the following Grover parameters:  $\rho=0.5$\si{\metre} and $d=1$\si{\metre}. The results of
calculation of diagonal and non-diagonal components of magnetic stiffness are as follows

\vspace*{1.0em}
\begin{footnotesize}
\begin{tabular}{lccc}
 \toprule
 &  BM, \si{\newton\per\metre}   &MIM, \si{\newton\per\metre}& GM, \si{\newton\per\metre} \\
   &  Eqs (\ref{eq:Sxx}), (\ref{eq:Syy}), (\ref{eq:Szz}) & Eqs (\ref{eq:Sqq NC}), (\ref{eq:Szz NC}) & Eqs (\ref{eq:Srho GF}), (\ref{eq:Sd GF}) \\
    \midrule
  $S_{xx}$ & $1.402100143236235\times10^{-7} $  & $1.402100143236235\times10^{-7} $& $1.402100143236236\times10^{-7} $  \\
   $S_{yy}$ & $2.118309158188127\times10^{-7} $ & $2.118309158188122\times10^{-7} $ &  $2.118309158188126\times10^{-7} $  \\
    $S_{zz}$ &  $-3.520409301424362\times10^{-7} $  & $-3.520409301424361\times10^{-7} $ & $-3.520409301424361\times10^{-7}$ \\
  \toprule
\end{tabular}
\end{footnotesize}

\vspace*{1.0em}
\begin{footnotesize}
\begin{tabular}{lccc}
 \toprule
 &  BM, \si{\newton\per\metre}   &MIM, \si{\newton\per\metre}& GM, \si{\newton\per\metre} \\
   &  Eqs (\ref{eq:Sxy}), (\ref{eq:Syz}), (\ref{eq:Sxz}) & Eqs (\ref{eq:Sqz NC}), (\ref{eq:Sxy NC}) & Eqs (\ref{eq:Sdrho GF})\\
    \midrule
  $S_{xy}$ &   $3.873036680241143\times10^{-141} $  & $-2.629904868295323\times10^{-22} $&NA  \\
   $S_{yz}$ & $9.040026778347652\times10^{-8} $ & $9.040026778347634\times10^{-8} $ &  $9.040026778347627\times10^{-8} $  \\
    $S_{xz}$ &   $-1.38528525192781\times10^{-140} $  & $9.028031637431708\times10^{-23} $ & NA \\
  \toprule
\end{tabular}
\end{footnotesize}

The component $S_{xx}$ in GM was restored form the condition (\ref{eq:sum condition}).

\subsubsection*{ Example 3}\label{sec:exmple3}
The coils having the same radii as in the previous examples and the center of the secondary coil is located at the point $x_c=0$\si{\metre}, $y_c=0$\si{\metre} and $z_c=0$\si{\metre}. Grover parameters are zeros.  The results of
calculation of diagonal and non-diagonal components of magnetic stiffness are as follows

\vspace*{1.0em}
\begin{footnotesize}
\begin{tabular}{lccc}
 \toprule
 &  BM, \si{\newton\per\metre}   &MIM, \si{\newton\per\metre}& GM, \si{\newton\per\metre} \\
   &  Eqs (\ref{eq:Sxx}), (\ref{eq:Syy}), (\ref{eq:Szz}) & Eqs (\ref{eq:Sqq NC}), (\ref{eq:Szz NC}) & Eqs (\ref{eq:Srho GF}), (\ref{eq:Sd GF}) \\
    \midrule
  $S_{xx}$ & $-6.367128613342259\times10^{-7} $  & $-6.367128613342232\times10^{-7} $& $-6.367128613342277\times10^{-7} $  \\
   $S_{yy}$ & $-6.367128613342259\times10^{-7} $ & $-6.36712861334223\times10^{-7} $ &   $-6.367128613342218\times10^{-7} $  \\
    $S_{zz}$ &  $1.273425722668452 \times10^{-6} $  & $1.273425722668452\times10^{-6} $ &  $1.273425722668449\times10^{-6} $ \\
  \toprule
\end{tabular}
\end{footnotesize}

\vspace*{1.0em}
\begin{footnotesize}
\begin{tabular}{lccc}
 \toprule
 &  BM, \si{\newton\per\metre}   &MIM, \si{\newton\per\metre}& GM, \si{\newton\per\metre} \\
   &  Eqs (\ref{eq:Sxy}), (\ref{eq:Syz}), (\ref{eq:Sxz}) & Eqs (\ref{eq:Sqz NC}), (\ref{eq:Sxy NC}) & Eqs (\ref{eq:Sdrho GF})\\
    \midrule
  $S_{xy}$ &   $-1.752675194540019\times10^{-141} $  & $-2.271727526158087\times10^{-22} $&NA  \\
   $S_{yz}$ & 0 & $0 $ &  0  \\
    $S_{xz}$ &0 & $0 $ & 0 \\
  \toprule
\end{tabular}
\end{footnotesize}

\begin{figure}[!t]
  \centering
  \includegraphics[width=2.5in]{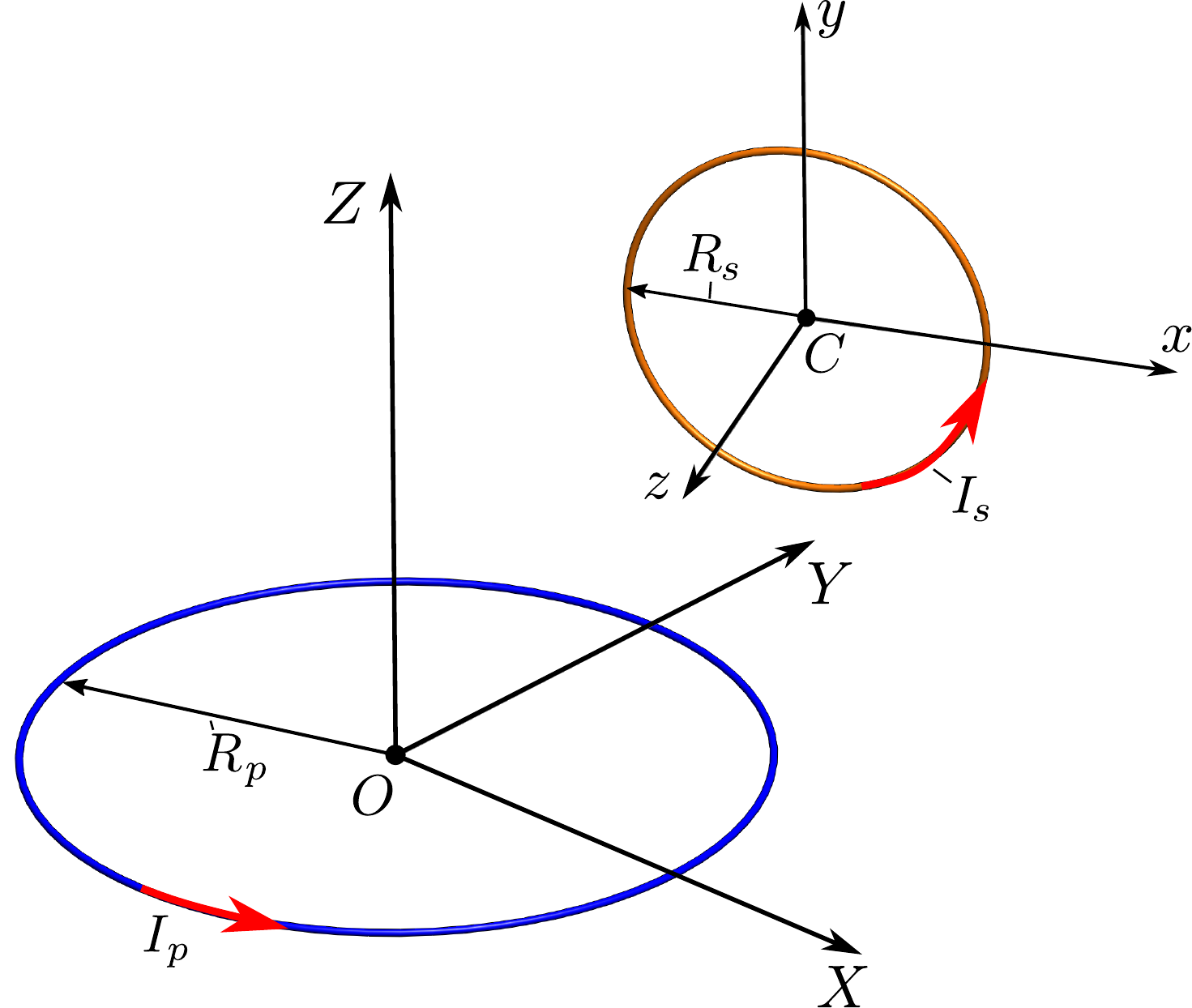}
  \caption{Geometrical scheme of mutually perpendicular current-carrying circular filaments (the second case: $\theta=\pi/2$).   }\label{fig:filaments perpendicular}
\end{figure}

\subsection{Magnetic stiffness between mutually perpendicular circular  filaments}
In this section, magnetic stiffness between mutually perpendicular current-carrying filaments is calculated. The general scheme is shown in Fig. \ref{fig:filaments   perpendicular}. For the mutual inductance method, it corresponds to the second case, when $\theta=\pi/2$ and the angular misalignment is only characterized by the angle $\eta$,  and the set of formulas (\ref{eq:Sqq SC})-(\ref{eq:Sxy SC}) is used.

\subsubsection*{ Example 4}
The two mutually perpendicular coils have the following radii, namely, $R_p = 0.2$ \si{\metre}, and the secondary one $R_s = 0.1$ \si{\metre}.  The center of the secondary coil is located at the origin as shown in Fig. \ref{fig:Example 4}. The angle $\eta$ is zero \si{\radian} for MIM. For BM, the angular misalignment is characterized by the following components of plane equation (\ref{eq:inclined plane}): $a=c=0$ and $b=1$.
The results of
calculation of diagonal and non-diagonal components of magnetic stiffness are as follows

\vspace*{1.0em}
\begin{footnotesize}
\begin{tabular}{lccc}
 \toprule
 &  BM, \si{\newton\per\metre}   &MIM, \si{\newton\per\metre}& GM, \si{\newton\per\metre} \\
   &  Eqs (\ref{eq:Sxx}), (\ref{eq:Syy}), (\ref{eq:Szz}) & Eqs (\ref{eq:Sqq SC}), (\ref{eq:Szz SC}) & Eqs (\ref{eq:Srho GF}), (\ref{eq:Sd GF}) \\
    \midrule
  $S_{xx}$ & 0  & 0& Not a Number (NaN)  \\
   $S_{yy}$ & 0 & 0 &   NaN  \\
    $S_{zz}$ &  0  & 0 &  NaN \\
  \toprule
\end{tabular}
\end{footnotesize}

\vspace*{1.0em}
\begin{footnotesize}
\begin{tabular}{lccc}
 \toprule
 &  BM, \si{\newton\per\metre}   &MIM, \si{\newton\per\metre}& GM, \si{\newton\per\metre} \\
   &  Eqs (\ref{eq:Sxy}), (\ref{eq:Syz}), (\ref{eq:Sxz}) & Eqs (\ref{eq:Sqz SC}), (\ref{eq:Sxy SC}) & Eqs (\ref{eq:Sdrho GF})\\
    \midrule
  $S_{xy}$ &   0  & 0&NA  \\
   $S_{yz}$ & $2.706560599934499\times10^{-6}  $& $2.706560599933974\times10^{-6}  $ & NaN  \\
    $S_{xz}$ &0 & $0 $ & NA \\
  \toprule
\end{tabular}
\end{footnotesize}

Note that for Grover's method it is the singular case.
\begin{figure*}[!t]
    \centering
     \subfigure[Example 4]
    {
    \centering
        \includegraphics[width=1.8in]{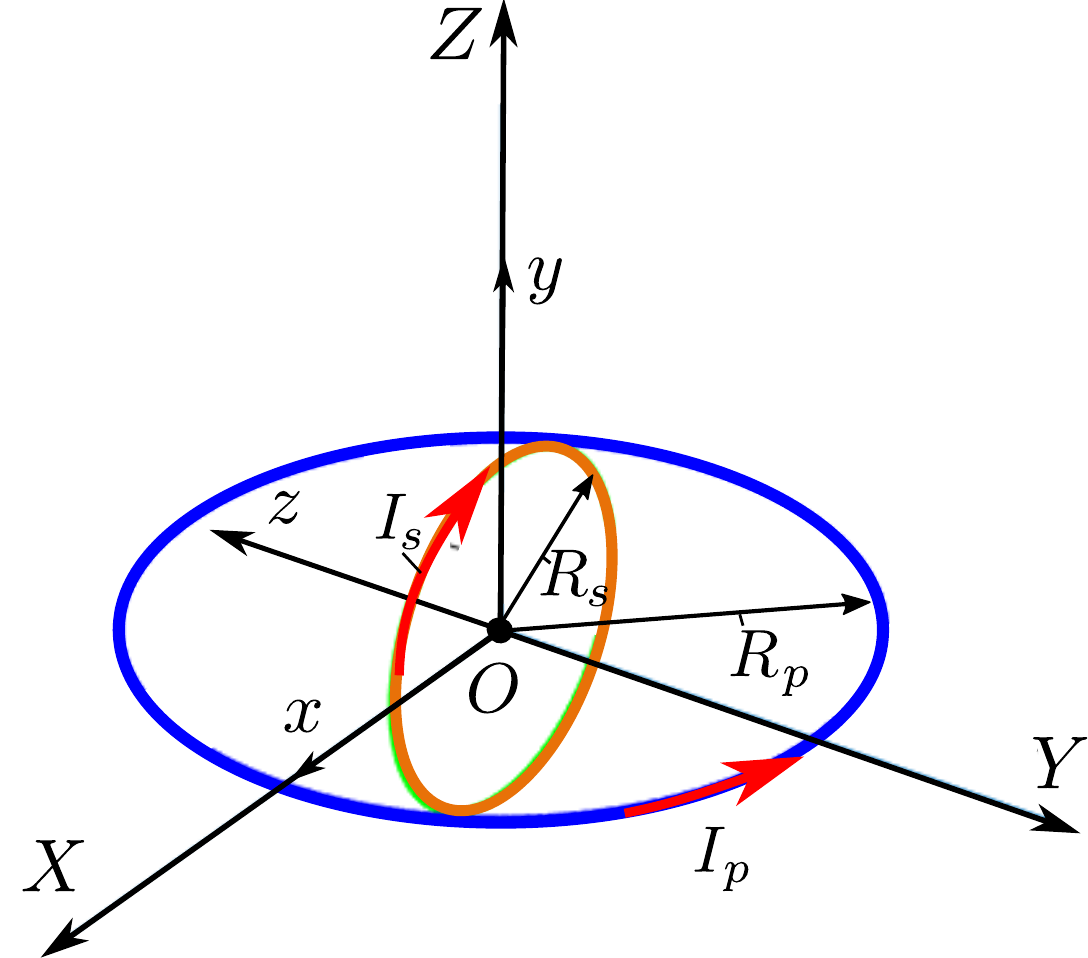}
       \label{fig:Example 4}
        }\quad
        \subfigure[Example 5 ]
    {
    \centering
        \includegraphics[width=1.8in]{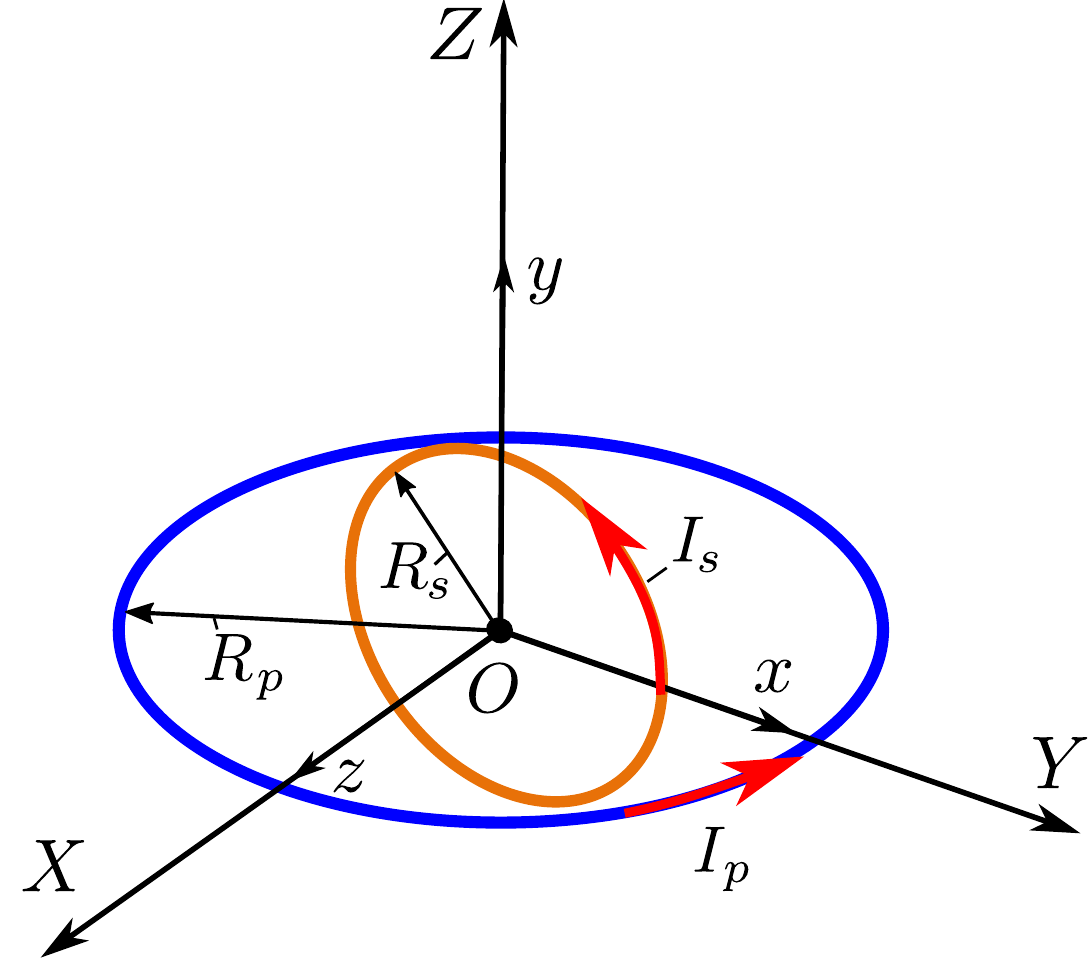}
       \label{fig:Example 5}
        }\quad
    \caption{Schemes of configuration of mutually perpendicular current-carrying coils.  }
    \label{fig:Example 4 and 5}
\end{figure*}

\subsubsection*{ Example 5}
 The same arrangement of coils as in example 4 is used, but the secondary coil is turned on the angle  $\eta=\pi/2$  \si{\radian} as shown in Fig. \ref{fig:Example 5}. For BM, the angular misalignment is characterized by the following components of plane equation (\ref{eq:inclined plane}): $b=c=0$.
 The results of
calculation are as follows

\vspace*{1.0em}
\begin{footnotesize}
\begin{tabular}{lccc}
 \toprule
 &  BM, \si{\newton\per\metre}   &MIM, \si{\newton\per\metre}& GM, \si{\newton\per\metre} \\
   &  Eqs (\ref{eq:Sxx}), (\ref{eq:Syy}), (\ref{eq:Szz}) & Eqs (\ref{eq:Sqq SC}), (\ref{eq:Szz SC}) & Eqs (\ref{eq:Srho GF}), (\ref{eq:Sd GF}) \\
    \midrule
  $S_{xx}$ & 0  & 0& Not a Number (NaN)  \\
   $S_{yy}$ & 0 & 0 &   NaN  \\
    $S_{zz}$ &  0  & 0 &  NaN \\
  \toprule
\end{tabular}
\end{footnotesize}

\vspace*{1.0em}
\begin{footnotesize}
\begin{tabular}{lccc}
 \toprule
 &  BM, \si{\newton\per\metre}   &MIM, \si{\newton\per\metre}& GM, \si{\newton\per\metre} \\
   &  Eqs (\ref{eq:Sxy}), (\ref{eq:Syz}), (\ref{eq:Sxz}) & Eqs (\ref{eq:Sqz SC}), (\ref{eq:Sxy SC}) & Eqs (\ref{eq:Sdrho GF})\\
    \midrule
  $S_{xy}$ &   0  & 0&NA \\
   $S_{yz}$ &0  & $1.657290387703741\times10^{-22}  $ & NaN  \\
    $S_{xz}$ &$-2.706560599933975\times10^{-6}  $  & $-2.706560599933974\times10^{-6}  $  & NA \\
  \toprule
\end{tabular}
\end{footnotesize}

Note that for Grover's method it is the singular case.

\subsubsection*{ Example 6}
The two mutually perpendicular coils have the following radii, namely, $R_p = 1.0$ \si{\metre}, and the secondary one $R_s = 0.5$ \si{\metre}.  The center of the secondary coil is located at the point $x_c=0$\si{\metre}, $y_c=2$\si{\metre} and $z_c=0$\si{\metre} . The angle $\eta$ is zero \si{\radian} for MIM. For BM, the angular misalignment is characterized by the following components of plane equation (\ref{eq:inclined plane}): $a=c=0$ and $b=1$.
The results of
calculation of diagonal and non-diagonal components of magnetic stiffness are as follows

\vspace*{1.0em}
\begin{footnotesize}
\begin{tabular}{lccc}
 \toprule
 &  BM, \si{\newton\per\metre}   &MIM, \si{\newton\per\metre}& GM, \si{\newton\per\metre} \\
   &  Eqs (\ref{eq:Sxx}), (\ref{eq:Syy}), (\ref{eq:Szz}) & Eqs (\ref{eq:Sqq SC}), (\ref{eq:Szz SC}) & Eqs (\ref{eq:Srho GF}), (\ref{eq:Sd GF}) \\
    \midrule
  $S_{xx}$ & 0  & 0&  $-3.68315244876568\times10^{-24}  $ \\
   $S_{yy}$ & $1.19859223988025\times10^{-143}  $  & 0 &   $8.191673408378001\times10^{-24}  $  \\
    $S_{zz}$ & $4.794368959521\times10^{-142}  $  & 0 &  $-4.508520959612321\times10^{-24}  $\\
  \toprule
\end{tabular}
\end{footnotesize}

\vspace*{1.0em}
\begin{footnotesize}
\begin{tabular}{lccc}
 \toprule
 &  BM, \si{\newton\per\metre}   &MIM, \si{\newton\per\metre}& GM, \si{\newton\per\metre} \\
   &  Eqs (\ref{eq:Sxy}), (\ref{eq:Syz}), (\ref{eq:Sxz}) & Eqs (\ref{eq:Sqz SC}), (\ref{eq:Sxy SC}) & Eqs (\ref{eq:Sdrho GF})\\
    \midrule
  $S_{xy}$ &  0   & 0&NA  \\
   $S_{yz}$ &$-1.368742764885786\times10^{-7}  $  & $-1.368742764885786\times10^{-7}  $ &$-1.368742764885786\times10^{-7}  $   \\
    $S_{xz}$ &0 & $5.151865880056137\times10^{-24}  $  & NA \\
  \toprule
\end{tabular}
\end{footnotesize}

\subsubsection*{ Example 7}
The two mutually perpendicular coils have the following radii, namely, $R_p = 1.0$ \si{\metre}, and the secondary one $R_s = 0.5$ \si{\metre}.  The center of the secondary coil is located at the point $x_c=0$\si{\metre}, $y_c=2$\si{\metre} and $z_c=3$\si{\metre} . The angle $\eta$ is zero \si{\radian} for MIM. For BM, the angular misalignment is characterized by the following components of plane equation (\ref{eq:inclined plane}): $a=c=0$ and $b=1$.
The results of
calculation of diagonal and non-diagonal components of magnetic stiffness are as follows

\vspace*{1.0em}
\begin{footnotesize}
\begin{tabular}{lccc}
 \toprule
 &  BM, \si{\newton\per\metre}   &MIM, \si{\newton\per\metre}& GM, \si{\newton\per\metre} \\
   &  Eqs (\ref{eq:Sxx}), (\ref{eq:Syy}), (\ref{eq:Szz}) & Eqs (\ref{eq:Sqq SC}), (\ref{eq:Szz SC}) & Eqs (\ref{eq:Srho GF}), (\ref{eq:Sd GF}) \\
    \midrule
  $S_{xx}$ & $-2.262682905005172\times10^{-9}  $  &  $-2.262682905005171\times10^{-9}  $ &  $-2.262682905005174\times10^{-9}  $ \\
   $S_{yy}$ & $-2.710919377082796\times10^{-9}  $ &  $-2.710919377082797\times10^{-9}  $  &   $-2.710919377082796\times10^{-9}  $  \\
    $S_{zz}$ &   $4.973602282087952\times10^{-9}  $ & $4.973602282087968\times10^{-9}  $  &  $4.97360228208797\times10^{-9}  $\\
  \toprule
\end{tabular}
\end{footnotesize}

\vspace*{1.0em}
\begin{footnotesize}
\begin{tabular}{lccc}
 \toprule
 &  BM, \si{\newton\per\metre}   &MIM, \si{\newton\per\metre}& GM, \si{\newton\per\metre} \\
   &  Eqs (\ref{eq:Sxy}), (\ref{eq:Syz}), (\ref{eq:Sxz}) & Eqs (\ref{eq:Sqz SC}), (\ref{eq:Sxy SC}) & Eqs (\ref{eq:Sdrho GF})\\
    \midrule
  $S_{xy}$ &  $-2.364186955151571\times10^{-143}  $   & $-2.347855544296417\times10^{-25}  $&NA  \\
   $S_{yz}$ &$3.101402573517489\times10^{-9}  $   &  $3.101402573517489\times10^{-9}  $   &$3.101402573517489\times10^{-9}  $   \\
    $S_{xz}$ & $7.984780962755199\times10^{-143}  $  & $8.433113791758558\times10^{-26}  $ & NA \\
  \toprule
\end{tabular}
\end{footnotesize}

\begin{figure}[!t]
  \centering
  \includegraphics[width=2.5in]{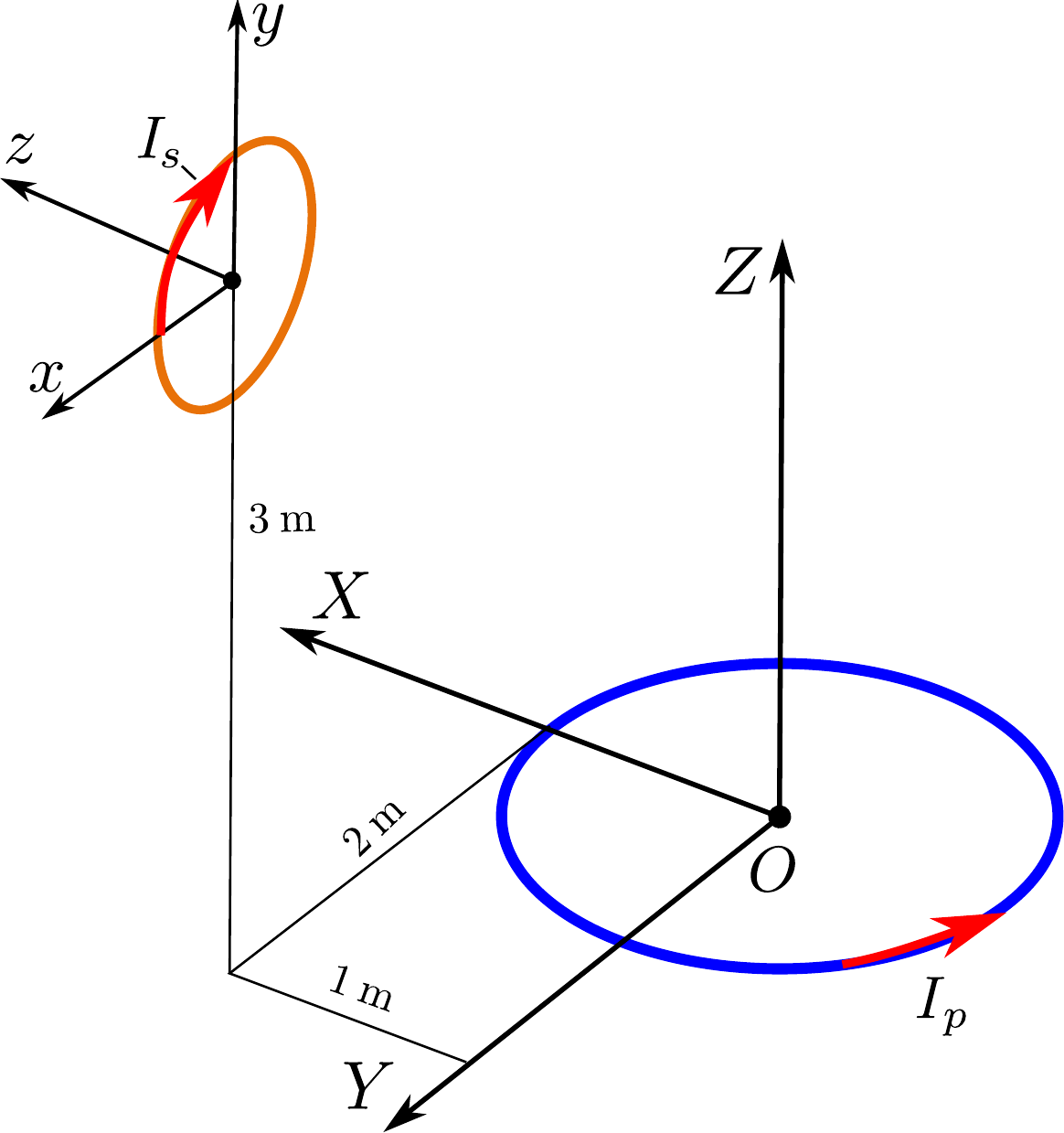}
  \caption{Geometrical scheme of mutually perpendicular current-carrying circular filaments for Example 8.   }\label{fig:Example8}
\end{figure}

\subsubsection*{ Example 8}
The two mutually perpendicular coils have the following radii, namely, $R_p = 1.0$ \si{\metre}, and the secondary one $R_s = 0.5$ \si{\metre}.  The center of the secondary coil is located at the point $x_c=1$\si{\metre}, $y_c=2$\si{\metre} and $z_c=3$\si{\metre}. The angle $\eta$ is $\pi/2$ \si{\radian} for MIM as shown in Fig. \ref{fig:Example8}. For BM, the angular misalignment is characterized by the following components of plane equation (\ref{eq:inclined plane}): $b=c=0$ and $a=1$.
The results of
calculation 
are as follows

\vspace*{1.0em}
\begin{footnotesize}
\begin{tabular}{lccc}
 \toprule
 &  BM, \si{\newton\per\metre}   &MIM, \si{\newton\per\metre}& GM, \si{\newton\per\metre} \\
   &  Eqs (\ref{eq:Sxx}), (\ref{eq:Syy}), (\ref{eq:Szz}) & Eqs (\ref{eq:Sqq SC}), (\ref{eq:Szz SC}) & Eqs (\ref{eq:Srho GF}), (\ref{eq:Sd GF}) \\
    \midrule
  $S_{xx}$ &   $2.444411106760408\times10^{-9}  $ &  $2.444411106760407\times10^{-9}  $ & NA \\
   $S_{yy}$ &  $-6.546286516635751\times10^{-10}  $  &  $-6.546286516635743\times10^{-10}  $  &   NA  \\
    $S_{zz}$ & $-1.789782455096833\times10^{-9}  $    & $-1.789782455096833\times10^{-9}  $  &  NA\\
  \toprule
\end{tabular}
\end{footnotesize}

\vspace*{1.0em}
\begin{footnotesize}
\begin{tabular}{lccc}
 \toprule
 &  BM, \si{\newton\per\metre}   &MIM, \si{\newton\per\metre}& GM, \si{\newton\per\metre} \\
   &  Eqs (\ref{eq:Sxy}), (\ref{eq:Syz}), (\ref{eq:Sxz}) & Eqs (\ref{eq:Sqz SC}), (\ref{eq:Sxy SC}) & Eqs (\ref{eq:Sdrho GF})\\
    \midrule
  $S_{xy}$ &  $1.042889962848133\times10^{-9}  $   & $1.042889962848133\times10^{-9}  $&NA  \\
   $S_{yz}$ &  $-2.190346410345056\times10^{-9}  $& $-2.190346410345057\times10^{-9}  $ &NA   \\
    $S_{xz}$ &$1.067688019471112\times10^{-9}  $ & $1.067688019471113\times10^{-9}  $ & NA \\
  \toprule
\end{tabular}
\end{footnotesize}

\subsection{Magnetic stiffness between circular  filaments arbitrarily positioned in the
space}

In this section, using the equation of inclined plane (\ref{eq:inclined plane})  for BM and its relationship with the Grover's angles (\ref{eq:relatioship bet angles and constants}) for MIM to define different angular misalignments of the secondary coil with respect to the primary one, a number of examples  with different arrangements of coils for calculation of magnetic stiffness are designed and considered below. The calculation is accompanied  by the evaluation of stiffness by means of Grover's formulas Eqs (\ref{eq:Srho GF}), (\ref{eq:Sd GF}) and  (\ref{eq:Sdrho GF}) when they are applicable.

\begin{figure}[!t]
  \centering
  \includegraphics[width=2.0in]{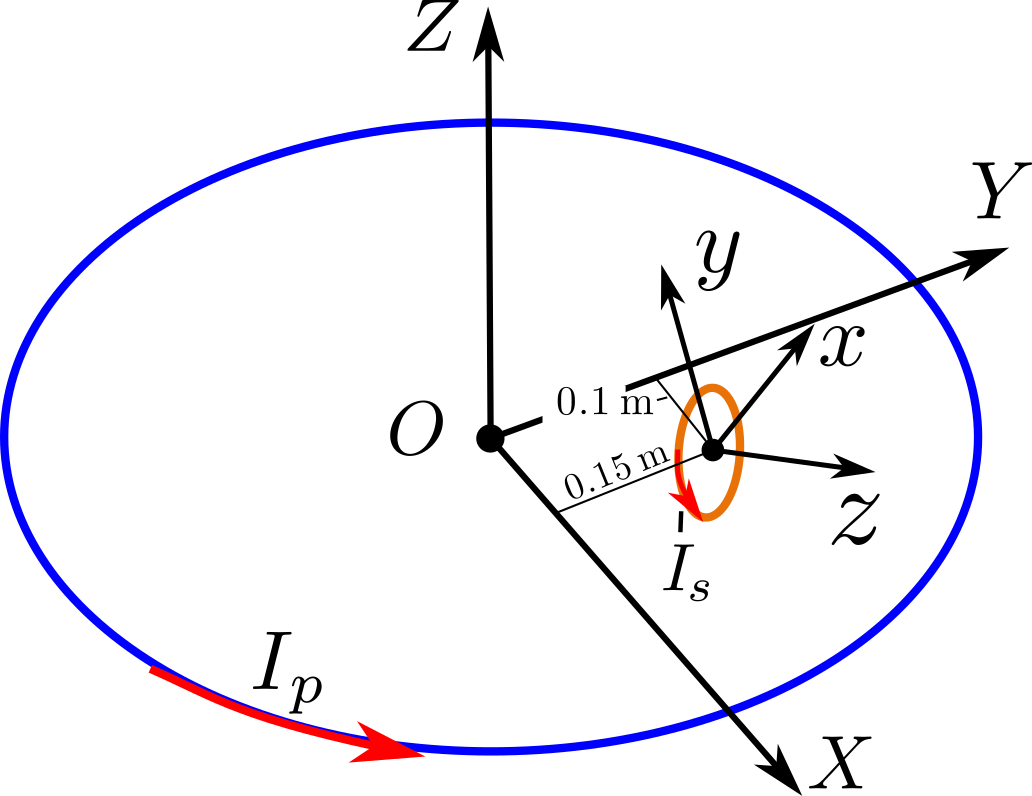}
  \caption{Geometrical scheme for Example 9.   }\label{fig:Example9}
\end{figure}
\subsubsection*{ Example 9}
The two coils have the following radii, namely, the primery one  $R_p = 0.4$ \si{\metre}, and the secondary one $R_s = 0.05$ \si{\metre}.  The center of the secondary coil is located at the point $x_c=0.1$\si{\metre}, $y_c=0.15$\si{\metre} and $z_c=0$\si{\metre}. For BM, the angular misalignment is characterized by the following  plane equation, namely, $3x+2y+z=0.6$. According to Eqs. (\ref{eq:relatioship bet angles and constants}), it corresponds to the angle $\eta =2.15879893034246$ \si{\radian} ($123.69006752598$\si{\degree})  and $\theta =1.30024656381632$ \si{\radian} ($74.498640433063$\si{\degree}) in notations of MIM. The coils' arrangement is  shown in Fig. \ref{fig:Example9}.
The results of
calculation 
are as follows

\vspace*{1.0em}
\begin{footnotesize}
\begin{tabular}{lccc}
 \toprule
 &  BM, \si{\newton\per\metre}   &MIM, \si{\newton\per\metre}& GM, \si{\newton\per\metre} \\
   &  Eqs (\ref{eq:Sxx}), (\ref{eq:Syy}), (\ref{eq:Szz}) & Eqs (\ref{eq:Sqq NC}), (\ref{eq:Szz NC})& Eqs (\ref{eq:Srho GF}), (\ref{eq:Sd GF}) \\
    \midrule
  $S_{xx}$ &  $-5.327433787498592\times10^{-8}  $ &  $-5.32743378749859\times10^{-8}  $ & NA \\
   $S_{yy}$ & $-7.00453721025121\times10^{-8}  $  &  $-7.004537210251179\times10^{-8}  $  &   NA  \\
    $S_{zz}$ & $1.23319709977498\times10^{-7}  $     & $1.23319709977498\times10^{-7}  $  &  NA\\
  \toprule
\end{tabular}
\end{footnotesize}

\vspace*{1.0em}
\begin{footnotesize}
\begin{tabular}{lccc}
 \toprule
 &  BM, \si{\newton\per\metre}   &MIM, \si{\newton\per\metre}& GM, \si{\newton\per\metre} \\
   &  Eqs (\ref{eq:Sxy}), (\ref{eq:Syz}), (\ref{eq:Sxz}) & Eqs (\ref{eq:Sqz NC}), (\ref{eq:Sxy NC}) &Eq (\ref{eq:Sdrho GF})\\
    \midrule
  $S_{xy}$ &  $-1.500101622143787\times10^{-8} $ & $-1.500101622143775\times10^{-8} $&NA  \\
   $S_{yz}$ &  $-2.121823074384375\times10^{-7}  $ & $-2.121823074384376\times10^{-7}  $ &  NA \\
    $S_{xz}$ &$-2.167771909796546\times10^{-7} $ & $-2.167771909796545\times10^{-7} $& NA \\
  \toprule
\end{tabular}
\end{footnotesize}

\begin{figure}[!t]
  \centering
  \includegraphics[width=2.0in]{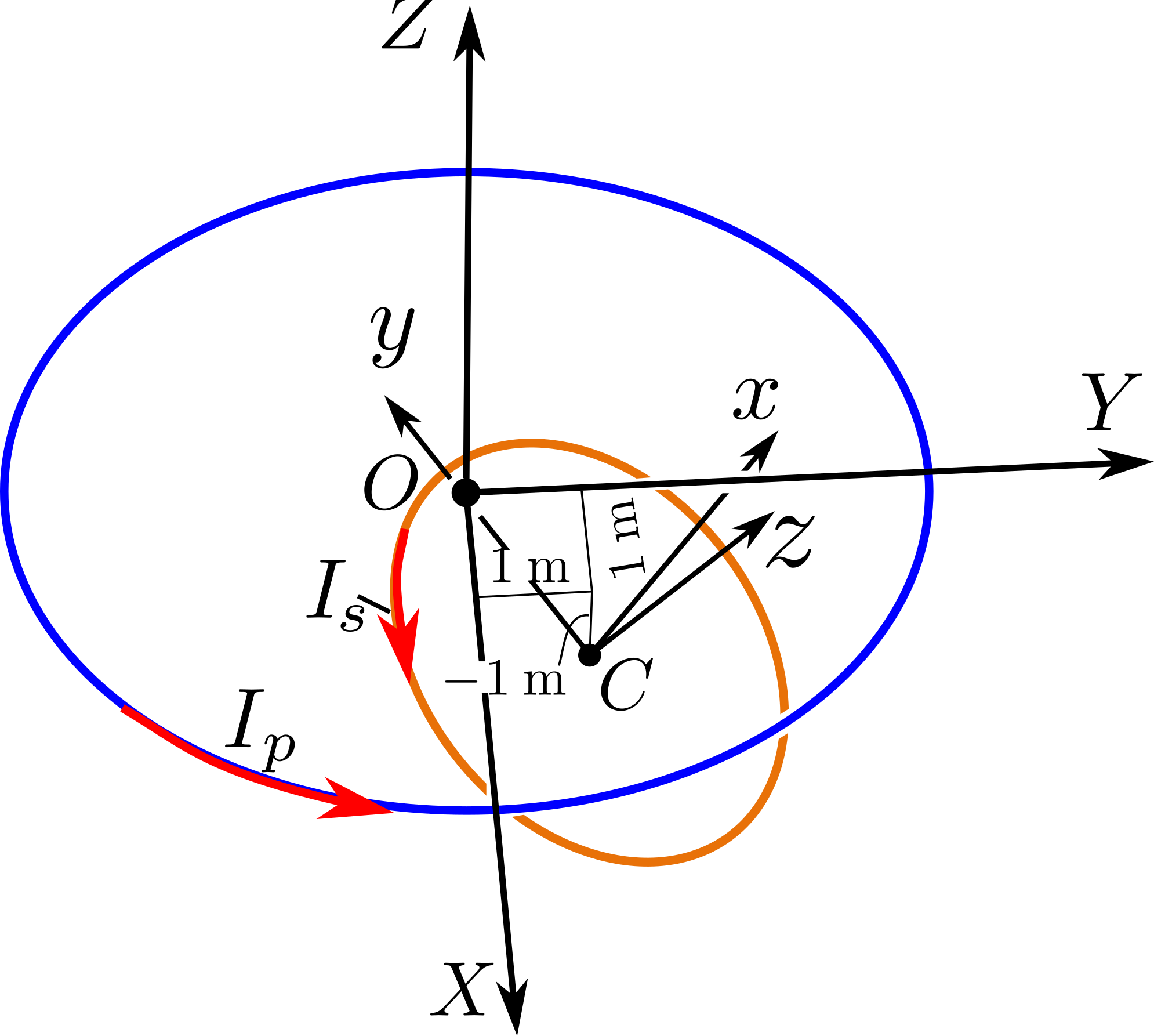}
  \caption{Geometrical scheme for Example 10.   }\label{fig:Example10}
\end{figure}
\subsubsection*{ Example 10}
The primery coil has a radius of  $R_p = 4$ \si{\metre}, and the secondary one has a radius of $R_s = 2$ \si{\metre}.  The center of the secondary coil is located at the point $x_c=1$\si{\metre}, $y_c=1$\si{\metre} and $z_c=-1$\si{\metre}. For BM, the angular misalignment is characterized by the following  plane equation, namely, $x+2y+3z=0$. According to Eqs. (\ref{eq:relatioship bet angles and constants}), the angles $\eta$ and $\theta$ are $2.67794504458899$ \si{\radian} ($153.434948822922
$\si{\degree})  and $0.640522312679424$ \si{\radian} ($36.6992252004899$\si{\degree}), respectively, in notations of MIM. The coils' arrangement is  shown in Fig. \ref{fig:Example10}.
The results of
calculation 
are as follows

\vspace*{1.0em}
\begin{footnotesize}
\begin{tabular}{lccc}
 \toprule
 &  BM, \si{\newton\per\metre}   &MIM, \si{\newton\per\metre}& GM, \si{\newton\per\metre} \\
   &  Eqs (\ref{eq:Sxx}), (\ref{eq:Syy}), (\ref{eq:Szz}) & Eqs (\ref{eq:Sqq NC}), (\ref{eq:Szz NC})& Eqs (\ref{eq:Srho GF}), (\ref{eq:Sd GF}) \\
    \midrule
  $S_{xx}$ &    $2.181662870952764\times10^{-8}  $  &  $2.181662870952587\times10^{-8}  $ & NA \\
   $S_{yy}$ &   $3.44815074198134\times10^{-8}  $  &  $3.448150741981526\times10^{-8}  $  &   NA  \\
    $S_{zz}$ & $-5.629813612934105\times10^{-8}  $     & $-5.629813612934103\times10^{-8}  $  &  NA\\
  \toprule
\end{tabular}
\end{footnotesize}

\vspace*{1.0em}
\begin{footnotesize}
\begin{tabular}{lccc}
 \toprule
 &  BM, \si{\newton\per\metre}   &MIM, \si{\newton\per\metre}& GM, \si{\newton\per\metre} \\
   &  Eqs (\ref{eq:Sxy}), (\ref{eq:Syz}), (\ref{eq:Sxz}) & Eqs (\ref{eq:Sqz NC}), (\ref{eq:Sxy NC}) & Eqs (\ref{eq:Sdrho GF})\\
    \midrule
  $S_{xy}$ &  $5.713524254486024\times10^{-8} $ & $5.713524254486125\times10^{-8} $&NA  \\
   $S_{yz}$ &  $-1.070006627660674\times10^{-7}  $ & $-1.07000662766067\times10^{-7}  $ &  NA \\
    $S_{xz}$ &$-1.105341440732599\times10^{-7} $ & $-1.1053414407326\times10^{-7} $& NA \\
  \toprule
\end{tabular}
\end{footnotesize}

\begin{figure}[!t]
  \centering
  \includegraphics[width=2.0in]{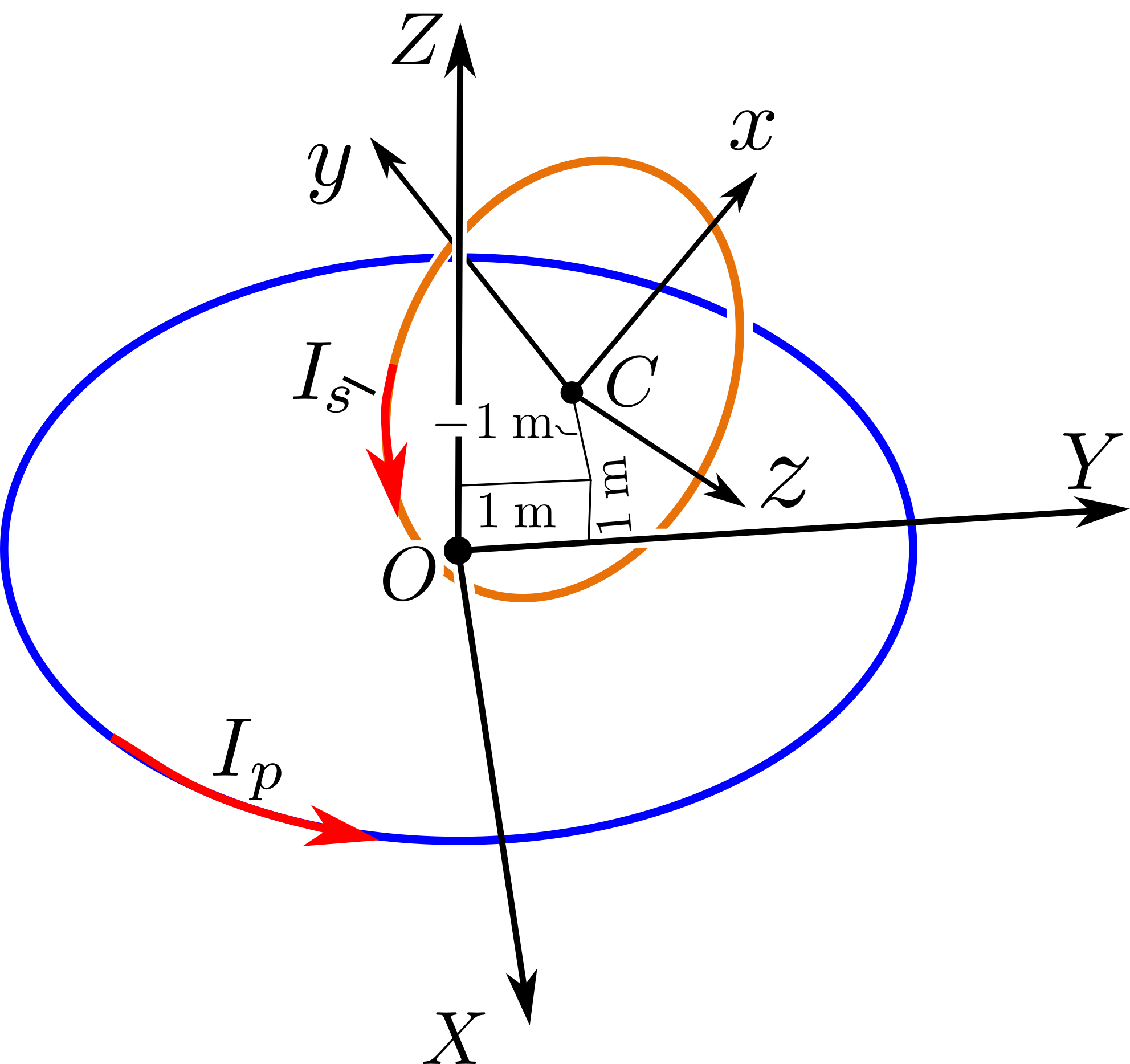}
  \caption{Geometrical scheme for Example 11.   }\label{fig:Example11}
\end{figure}
\subsubsection*{ Example 11}
The radii of the coils are the same as in the previous example 10,  but the center of the secondary coil is located at the point $x_c=-1$\si{\metre}, $y_c=1$\si{\metre} and $z_c=1$\si{\metre}. The angular misalignment for BM  is characterized by the following  plane equation, namely, $3x+2y+1z=0$. According to Eqs. (\ref{eq:relatioship bet angles and constants}), the angles $\eta$ and $\theta$ are $2.15879893034246$ \si{\radian} ($123.69006752598
$\si{\degree})  and $1.30024656381632$ \si{\radian} ($74.498640433063$\si{\degree}), respectively, in notations of MIM. The coils' arrangement is  shown in Fig. \ref{fig:Example11}.
The results of
calculation 
are as follows

\vspace*{1.0em}
\begin{footnotesize}
\begin{tabular}{lccc}
 \toprule
 &  BM, \si{\newton\per\metre}   &MIM, \si{\newton\per\metre}& GM, \si{\newton\per\metre} \\
   &  Eqs (\ref{eq:Sxx}), (\ref{eq:Syy}), (\ref{eq:Szz}) & Eqs (\ref{eq:Sqq NC}), (\ref{eq:Szz NC})& Eqs (\ref{eq:Srho GF}), (\ref{eq:Sd GF}) \\
    \midrule
  $S_{xx}$ &    $2.944440845967626\times10^{-8}  $  &  $2.944440845966999\times10^{-8}  $ & NA \\
   $S_{yy}$ &   $-5.401491525386883\times10^{-8}  $  &  $-5.401491525387853\times10^{-8}  $  &   NA  \\
    $S_{zz}$ & $2.457050679419257\times10^{-8}  $     & $2.457050679419274\times10^{-8}  $  &  NA\\
  \toprule
\end{tabular}
\end{footnotesize}

\vspace*{1.0em}
\begin{footnotesize}
\begin{tabular}{lccc}
 \toprule
 &  BM, \si{\newton\per\metre}   &MIM, \si{\newton\per\metre}& GM, \si{\newton\per\metre} \\
   &  Eqs (\ref{eq:Sxy}), (\ref{eq:Syz}), (\ref{eq:Sxz}) & Eqs (\ref{eq:Sqz NC}), (\ref{eq:Sxy NC}) & Eqs (\ref{eq:Sdrho GF})\\
    \midrule
  $S_{xy}$ &  $-7.826596371018449\times10^{-9} $ & $-7.826596371018387\times10^{-9} $&NA  \\
   $S_{yz}$ &  $-2.426184704880834\times10^{-8}  $ & $-2.426184704880832\times10^{-8}  $ &  NA \\
    $S_{xz}$ &$-1.147221843342089\times10^{-7} $ & $-1.14722184334209\times10^{-7} $& NA \\
  \toprule
\end{tabular}
\end{footnotesize}

\begin{figure}[!t]
  \centering
  \includegraphics[width=2.0in]{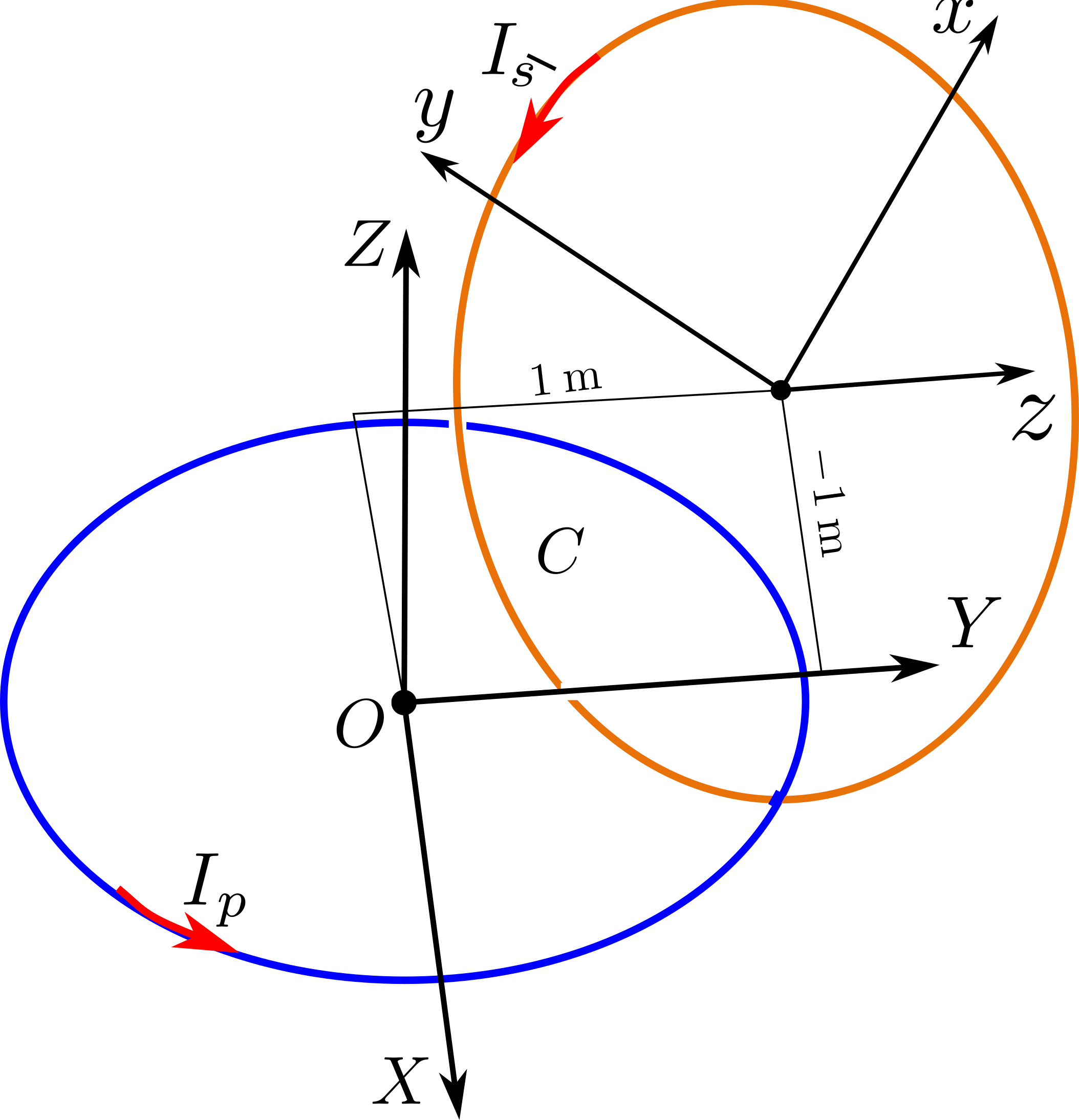}
  \caption{Geometrical scheme for Example 12: the linked coils.   }\label{fig:Example12}
\end{figure}
\subsubsection*{ Example 12}
The radii of each coil are the same and equal to  $1$ \si{\metre}.   
The center of the secondary coil is located at the point $x_c=-1$\si{\metre}, $y_c=1$\si{\metre} and $z_c=0$\si{\metre}. The angular misalignment for BM  is characterized by the following  plane equation, namely, $x+y+z=0$. According to Eqs. (\ref{eq:relatioship bet angles and constants}), the angles $\eta$ and $\theta$ are $2.35619449019234$ \si{\radian} ($135
$\si{\degree})  and $0.955316618124509$ \si{\radian} ($54.7356103172453$\si{\degree}), respectively, in notations of MIM. The arrangement of the linked coils is  shown in Fig. \ref{fig:Example12}.
The results of
calculation 
are as follows

\vspace*{1.0em}
\begin{footnotesize}
\begin{tabular}{lccc}
 \toprule
 &  BM, \si{\newton\per\metre}   &MIM, \si{\newton\per\metre}& GM, \si{\newton\per\metre} \\
   &  Eqs (\ref{eq:Sxx}), (\ref{eq:Syy}), (\ref{eq:Szz}) & Eqs (\ref{eq:Sqq NC}), (\ref{eq:Szz NC})& Eqs (\ref{eq:Srho GF}), (\ref{eq:Sd GF}) \\
    \midrule
  $S_{xx}$ &    $1.046477792966786\times10^{-7}  $  &  $1.046477792966789\times10^{-7}  $ & NaN \\
   $S_{yy}$ &   $1.046477792966786\times10^{-7}  $  &  $1.046477792966785\times10^{-7}  $  &   NaN  \\
    $S_{zz}$ & $-2.092955585933573\times10^{-8}  $     & $-2.092955585933573\times10^{-7}  $  &  NaN\\
  \toprule
\end{tabular}
\end{footnotesize}

\vspace*{1.0em}
\begin{footnotesize}
\begin{tabular}{lccc}
 \toprule
 &  BM, \si{\newton\per\metre}   &MIM, \si{\newton\per\metre}& GM, \si{\newton\per\metre} \\
   &  Eqs (\ref{eq:Sxy}), (\ref{eq:Syz}), (\ref{eq:Sxz}) & Eqs (\ref{eq:Sqz NC}), (\ref{eq:Sxy NC}) & Eqs (\ref{eq:Sdrho GF})\\
    \midrule
  $S_{xy}$ &  $2.014629707091004\times10^{-7} $ & $2.014629707091005\times10^{-7} $&NA \\
   $S_{yz}$ &  $ -2.577031542995681\times10^{-7}  $ & $-2.577031542995682\times10^{-7}  $ &  NaN \\
    $S_{xz}$ &$-2.577031542995681\times10^{-7} $ & $-2.577031542995681\times10^{-7} $& NA\\
  \toprule
\end{tabular}
\end{footnotesize}

\subsubsection*{ Example 13}
The radii of  coils  and their angular orienation with respect to each other are the same as in Example 12.  
The center of the secondary coil is located at the point $x_c=0$\si{\metre}, $y_c=1$\si{\metre} and $z_c=-1$\si{\metre}.
The results of
calculation 
are as follows

\vspace*{1.0em}
\begin{footnotesize}
\begin{tabular}{lccc}
 \toprule
 &  BM, \si{\newton\per\metre}   &MIM, \si{\newton\per\metre}& GM, \si{\newton\per\metre} \\
   &  Eqs (\ref{eq:Sxx}), (\ref{eq:Syy}), (\ref{eq:Szz}) & Eqs (\ref{eq:Sqq NC}), (\ref{eq:Szz NC})& Eqs (\ref{eq:Srho GF}), (\ref{eq:Sd GF}) \\
    \midrule
  $S_{xx}$ &    $ 2.986090969216481\times10^{-7}  $  &  $2.986090969216481\times10^{-7}  $ &$2.986090969216485\times10^{-7}  $\\
   $S_{yy}$ &   $-6.600031210636617\times10^{-8}  $  &  $-6.600031210636671\times10^{-8}  $  &     $-6.600031210636627\times10^{-8}  $    \\
    $S_{zz}$ & $-2.326087848152819\times10^{-8}  $     & $-2.326087848152818\times10^{-7}  $  &   $-2.326087848152822\times10^{-7}  $\\
  \toprule
\end{tabular}
\end{footnotesize}

\vspace*{1.0em}
\begin{footnotesize}
\begin{tabular}{lccc}
 \toprule
 &  BM, \si{\newton\per\metre}   &MIM, \si{\newton\per\metre}& GM, \si{\newton\per\metre} \\
   &  Eqs (\ref{eq:Sxy}), (\ref{eq:Syz}), (\ref{eq:Sxz}) & Eqs (\ref{eq:Sqz NC}), (\ref{eq:Sxy NC}) & Eqs (\ref{eq:Sdrho GF})\\
    \midrule
  $S_{xy}$ &  $-3.199237389569022\times10^{-8} $ & $-3.199237389569028\times10^{-8} $&NA \\
   $S_{yz}$ &  $ 8.115657627384119\times10^{-7}  $ & $8.115657627384113\times10^{-7}  $ &  $8.11565762738412\times10^{-8}  $ \\
    $S_{xz}$ &$2.749069587302089\times10^{-7} $ & $2.749069587302088\times10^{-7} $& NA\\
  \toprule
\end{tabular}
\end{footnotesize}

\begin{figure}[!t]
  \centering
  \includegraphics[width=2.0in]{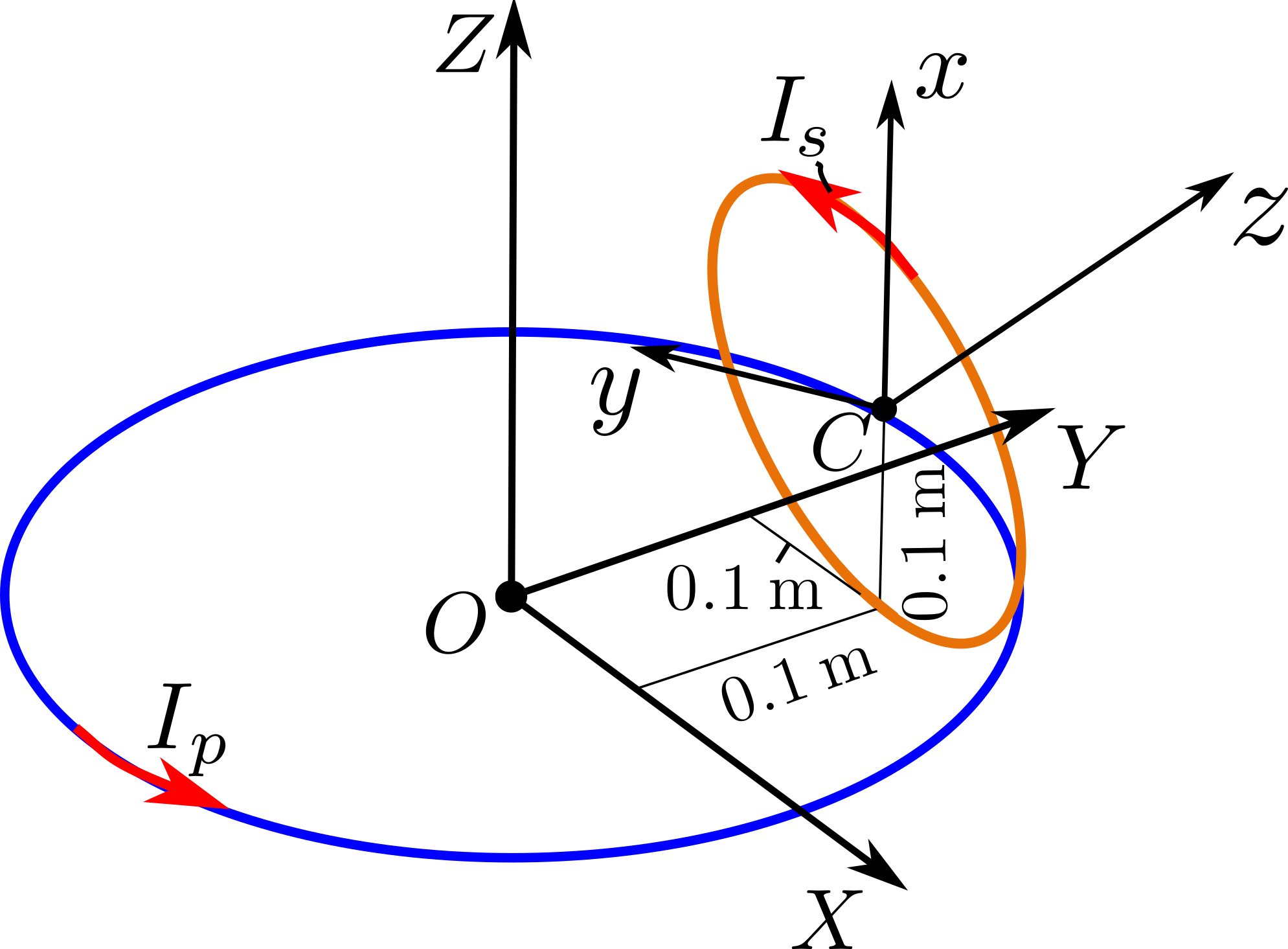}
  \caption{Geometrical scheme for Example 14.   }\label{fig:Example14}
\end{figure}
\subsubsection*{ Example 14}
The primary circular filament has a radius of $R_p = 0.2$ \si{\metre}, while the secondary one has $R_s = 0.1$ \si{\metre}.
The centre of secondary coil is located at the point $C$ having the following coordinates $x_c=0.1$\si{\metre}, $y_c=0.1$\si{\metre} and $z_c=0.1$\si{\metre}.
The angular misalignment for BM  is defined by the following  plane equation, namely, $x+y+z=0.3$. According to Eqs. (\ref{eq:relatioship bet angles and constants}), the angles $\eta$ and $\theta$ are $2.35619449019234$ \si{\radian} ($135
$\si{\degree})  and $0.955316618124509$ \si{\radian} ($54.7356103172453$\si{\degree}), respectively, in notations of MIM. The arrangement of the  coils is  shown in Fig. \ref{fig:Example14}.
The results of
calculation 
are as follows

\vspace*{1.0em}
\begin{footnotesize}
\begin{tabular}{lccc}
 \toprule
 &  BM, \si{\newton\per\metre}   &MIM, \si{\newton\per\metre}& GM, \si{\newton\per\metre} \\
   &  Eqs (\ref{eq:Sxx}), (\ref{eq:Syy}), (\ref{eq:Szz}) & Eqs (\ref{eq:Sqq NC}), (\ref{eq:Szz NC})& Eqs (\ref{eq:Srho GF}), (\ref{eq:Sd GF}) \\
    \midrule
  $S_{xx}$ &    $2.868431152918931\times10^{-5}  $  &   $2.868431152918925\times10^{-5}  $ & NA \\
   $S_{yy}$ &   $2.868431152918931\times10^{-5}  $  &  $2.868431152918923\times10^{-5}  $  &   NA  \\
    $S_{zz}$ & $-5.736862305837862\times10^{-5}  $     & $-5.736862305837814\times10^{-5}  $  &  NA\\
  \toprule
\end{tabular}
\end{footnotesize}

\vspace*{1.0em}
\begin{footnotesize}
\begin{tabular}{lccc}
 \toprule
 &  BM, \si{\newton\per\metre}   &MIM, \si{\newton\per\metre}& GM, \si{\newton\per\metre} \\
   &  Eqs (\ref{eq:Sxy}), (\ref{eq:Syz}), (\ref{eq:Sxz}) & Eqs (\ref{eq:Sqz NC}), (\ref{eq:Sxy NC}) & Eqs (\ref{eq:Sdrho GF})\\
    \midrule
  $S_{xy}$ &  $2.397140500000452 \times10^{-5} $ & $2.397140500000446\times10^{-5} $&NA \\
   $S_{yz}$ &  $ 1.724024033513611\times10^{-6}  $ & $1.724024033513529\times10^{-6}  $ &  NA \\
    $S_{xz}$ &$ 1.724024033513611\times10^{-6} $ & $1.724024033513543\times10^{-6} $& NA\\
  \toprule
\end{tabular}
\end{footnotesize}

\section{Conclusion}

 In this article,
  sets of analytical formulas for calculation of nine components of magnetic stiffness of corresponding force arising between two current-carrying circular filaments arbitrarily oriented in the space have been derived by using Babic's method and the method of mutual inductance (Kalantarov-Zeitlin’s method).  Formulas are presented  through integral
expressions, whose kernel function is expressed in terms of the elliptic integrals of the first and second kinds.
Also,  the additional set of  expressions for calculation of components of  magnetic stiffness by means of differentiation of Grover’s formula  with respect to appropriate coordinates has been obatined. Grover's method provides the most simplest approach for calculation of magnetic stiffness, however the calculation is constrained by  four components only, namely, two diaganal  and two non-diaganal components   instead of nine ones. Also, the GM suffers from singular cases  shown, for instance, in Examples 4, 5 and 12, which limit the applicability of the method. In  opposite to the GM, the set of formulas  (\ref{eq:Sxx})-(\ref{eq:Szz}), (\ref{eq:Sqq NC})-(\ref{eq:Sxy NC}) and (\ref{eq:Sqq SC})-(\ref{eq:Sxy SC}) is deduced by BM and MIM, respectively, is universally applicable for calculation of the magnetic stiffness and covers all possible arrangements between two current-carrying circular filaments .
The derived sets of formulas were mutually validated and results of calculation of components of magnetic stiffness agree well to each other.

The set of formulas (\ref{eq:Sqq NC})-(\ref{eq:Sxy NC}) and (\ref{eq:Sqq SC})-(\ref{eq:Sxy SC}) obtained by means of MIM and the set of expressions (\ref{eq:Srho GF})-(\ref{eq:Sdrho GF}) obtained by means of GM were  programmed by using the Matlab language. The Matlab files with the implemented formulas are available as supplementary materials to this article.

\section*{Acknowledgment}
Kirill Poletkin acknowledges with thanks the support from German Research Foundation (Grant KO 1883/37-1) under the priority programme SPP 2206.

\appendix

\section{Stiffness calculation. Grover's method (GM) \cite[page 207, Eq. (179)]{Grover2004}}
\label{app:Grover}
\begin{figure}[!t]
  \centering
  \includegraphics[width=2.5in]{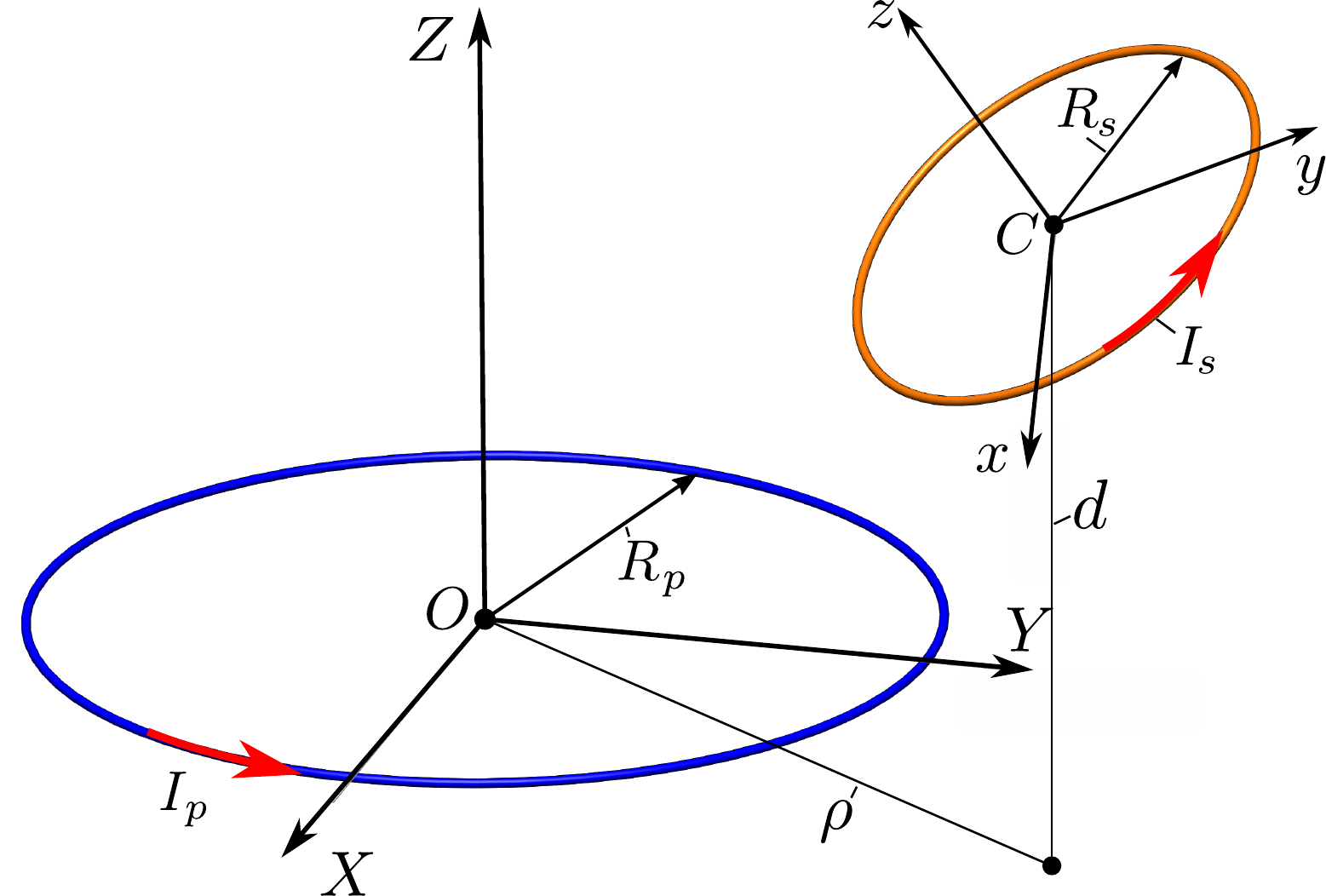}
  \caption{Geometrical scheme of current-carrying circular filaments arbitrarily oriented in the space: the Grover notations.   }\label{fig:filaments Grover notation}
\end{figure}

According to Grover's notations, the linear misalignment of the centre of the secondary circle is characterised by two parameters, namely, $d=z_c$ and $\rho=\sqrt{x_c^2+y_c^2}$ as shown in Figure \ref{fig:filaments Grover notation}. Besides that the angular misalignment is defined in accordance with the first manner as shown in Fig. \ref{fig:Grover angles}, but keeping the original Grover's notation the angle, $\eta$, is replaced by  $\psi$. In absence of the angular misalignment, the CF $xyz$ assigned to the secondary circle is oriented  in the following way. The $z$-axis is directed upward along the $d$-line, while the $y$-axis is parallel to the $\rho$-line and directed in continuation of the $\rho$-line. Then adopting the above considered notations, Grover's formula for calculation of mutual inductance between two circular filaments can be written as
\begin{equation}\label{eq:GROVER FORMULA}
  M=\frac{\mu_0\sqrt{R_pR_s}}{2\pi}\int_{0}^{2\pi} U\cdot\Psi(k)d\varphi,
\end{equation}
where
\begin{equation}\label{eq:U Grover}
  U=U({\gamma},\theta,\psi)=\frac{R({\gamma},\theta,\psi)}{{V}^{1.5}}=\frac{\cos\theta-\gamma(\cos\psi\cos\varphi-\sin\psi\cos\theta\sin\varphi)}{{V}^{1.5}},
\end{equation}
\begin{equation}\label{eq:V Grover}
V=V({\gamma},\theta,\psi)=\sqrt{1-{\cos(\varphi)}^2{\sin(\theta)}^2+2\gamma(\sin\psi\sin\varphi-\cos\varphi\cos\psi\cos\theta)+\gamma^2},
\end{equation}
\begin{equation}\label{eq:Psi Grover}
   \Psi(k)=\frac{2}{k}\left[\left(1-\frac{k^2}{2}\right)K(k)-E(k)\right],
\end{equation}
\begin{equation}\label{eq:k Grover}
\begin{array}{l}
   {\displaystyle k^2=k^2({\gamma},\Delta,\theta,\psi)=\frac{4\alpha{V}}{(\alpha{V}+1)^2+z^2},
}\\
 {\displaystyle \alpha=R_s/R_p,\;   \Delta=d/R_p,\; \gamma=\rho/R_s},\; z={\Delta}-{\alpha}\sin\theta\cos\varphi.
\end{array}
\end{equation}
The kernel of formula (\ref{eq:GROVER FORMULA}) is
\begin{equation}\label{eq:kernel NF}
  \mathrm{Kr}= U\cdot\Psi(k).
\end{equation}
Accounting for (\ref{eq:U Grover}), (\ref{eq:V Grover}), (\ref{eq:Psi Grover}) and (\ref{eq:k Grover}), the second $\rho$-derivative of the kernel has a  similar structure to Eq. (\ref{eq:second g-der of kernel SC}) and becomes as follows
\begin{equation}\label{eq:second rho-der of kernel Grover}
 \begin{array}{c}
   {\displaystyle  \frac{\partial^2\mathrm{Kr}}{\partial \rho^2}=\frac{\partial^2\mathrm{Kr}}{\partial \gamma^2}\frac{1}{R_s^2}=\frac{1}{R_s^2}\cdot\left[\frac{\partial^2 U}{\partial  \gamma^2} \cdot\Psi(k)+2\frac{\partial U}{\partial  \gamma}\cdot\frac{d \Psi(k)}{d k} \cdot\frac{\partial k}{\partial  \gamma}\right.} \\
   {\displaystyle \left. +U\left(\frac{d^2 \Psi(k)}{d k^2} \cdot\left(\frac{\partial k}{\partial  \gamma}\right)^2+\frac{d \Psi(k)}{d k} \cdot\frac{\partial^2 k}{\partial  \gamma^2} \right)\right],}
 \end{array}
\end{equation}
where
\begin{equation}\label{eq:dU-gamma}
 \begin{array}{l}
  {\displaystyle  \frac{\partial U}{\partial \gamma}=\frac{J}{V^{2.5}}=\left({\displaystyle\frac{\partial R}{\partial \gamma}}\cdot V-1.5\cdot R\cdot\frac{\partial V}{\partial \gamma}\right)\bigg/{{V}^{2.5}},}\\
  {\displaystyle  \frac{\partial^2 U}{\partial \gamma^2}=\frac{{\displaystyle\frac{\partial J}{\partial \gamma}\cdot V-2.5\cdot J\cdot\frac{\partial V}{\partial \gamma}}}{{{V}^{3.5}}},}\\
  {\displaystyle \frac{\partial J}{\partial \gamma}={\displaystyle-0.5\frac{\partial R}{\partial \gamma}\cdot \frac{\partial V}{\partial \gamma}}-1.5\cdot R\cdot\frac{\partial^2 V}{\partial \gamma^2}},\\
   {\displaystyle \frac{\partial R}{\partial \gamma}=-\cos\psi\cos\varphi+\sin\psi\cos\theta\sin\varphi},\\
  {\displaystyle \frac{\partial V}{\partial \gamma}=\frac{\sin\psi\sin\varphi-\cos\varphi\cos\psi\cos\theta+\gamma}{V}},\\
 {\displaystyle \frac{\partial^2 V}{\partial \gamma^2}=\frac{V-\left(\sin\psi\sin\varphi-\cos\varphi\cos\psi\cos\theta+\gamma\right) {\displaystyle\frac{\partial V}{\partial \gamma}}}{V^2}},
   \end{array}
\end{equation}
\begin{equation}\label{eq:dk-gamma}
  \begin{array}{l}
     {\displaystyle \frac{\partial k}{\partial \gamma}=\frac{G}{H}\cdot\alpha\frac{\partial V}{\partial \gamma},
} \\
 {\displaystyle \frac{\partial^2 k}{\partial \gamma^2}=\frac{ {\displaystyle\frac{\partial G}{\partial \gamma}H-A\frac{\partial H}{\partial \gamma}}}{H^2}\cdot\alpha\frac{\partial V}{\partial \gamma}+\frac{G}{H}\cdot\alpha\frac{\partial^2 V}{\partial \gamma^2},
}\\
  {\displaystyle G=2/k-k(\alpha{V}+1),\;  H=(\alpha{V}+1)^2+
  {z}^2,
}  \\
 {\displaystyle \frac{\partial G}{\partial \gamma}=-\left[2/k^2+\alpha{V}+1\right]\frac{\partial k}{\partial \gamma}-k\cdot\alpha\frac{\partial V}{\partial \gamma}, \frac{\partial H}{\partial \gamma}=2(\alpha V+1)\alpha\frac{\partial V}{\partial \gamma}.
}
  \end{array}
\end{equation}
Note that the first and second derivatives of function $\Psi$ with respect to $k$ are equal to ${\displaystyle 2\frac{d \Phi}{d k}}$ and ${\displaystyle 2\frac{d^2 \Phi}{d k^2}}$, respectively, where the  first and second derivatives of $\Phi$ with respect to $k$ are defined by Eqs. (\ref{eq:dPhi}).

 The second $d$-derivative of the kernel is
 \begin{equation}\label{eq:second z-der of kernel Grover}
   {\displaystyle  \frac{\partial^2\mathrm{Kr}}{\partial d^2}=\frac{\partial^2\mathrm{Kr}}{\partial \Delta^2}\frac{1}{R_p^2}=\frac{1}{R_p^2}\cdot U\left[\frac{d^2 \Psi(k)}{d k^2} \cdot\left(\frac{\partial k}{\partial \Delta}\right)^2+\frac{d \Psi(k)}{d k} \cdot\frac{\partial^2 k}{\partial \Delta^2} \right],}
\end{equation}
where
\begin{equation}\label{eq:dk-d}
  \begin{array}{l}
     {\displaystyle \frac{\partial k}{\partial \Delta}=\frac{G}{H},\; \frac{\partial^2 k}{\partial \Delta^2}=\frac{ {\displaystyle\frac{\partial G}{\partial \Delta}H-A\frac{\partial H}{\partial \Delta}}}{H^2},
} \\
  {\displaystyle G=-k\cdot z,\;  H=(\alpha{V}+1)^2+
  {z}^2,
}  \\
 {\displaystyle \frac{\partial G}{\partial \Delta}=-\frac{\partial k}{\partial \Delta}\cdot z-k, \frac{\partial H}{\partial \Delta}=2z.
}
  \end{array}
\end{equation}

Accounting for the property
\begin{equation}\label{eq:second rhod- der of kernel Grover}
 {\displaystyle  \frac{\partial^2\mathrm{Kr}}{\partial \rho\partial d}=\frac{\partial^2\mathrm{Kr}}{\partial d\partial \rho }, }
\end{equation}
the second derivative of the kernel with respect of variables $d$ and $\rho$ can be written as
\begin{equation}\label{eq:second drho-der of kernel Grover}
 \begin{array}{c}
   {\displaystyle  \frac{\partial^2\mathrm{Kr}}{\partial \rho \partial d}=\frac{\partial^2\mathrm{Kr}}{\partial \gamma \partial \Delta}\frac{1}{R_sR_p}=\frac{1}{R_sR_p}\cdot\left[\frac{\partial U}{\partial  \gamma } \cdot\frac{d \Psi(k)}{d k}\cdot\frac{\partial k}{\partial  \Delta}\right.} \\
   {\displaystyle \left. +U\left(\frac{d^2 \Psi(k)}{d k^2} \cdot\frac{\partial k}{\partial  \gamma}\cdot\frac{\partial k}{\partial  \Delta}+\frac{d \Psi(k)}{d k} \cdot\frac{\partial^2 k}{\partial  \gamma \partial \Delta} \right)\right],}
 \end{array}
\end{equation}
where
\begin{equation}\label{eq:dk-drho}
  \begin{array}{l}
     {\displaystyle  \frac{\partial^2 k}{\partial \gamma \partial \Delta}=\frac{ {\displaystyle\frac{\partial G}{\partial \gamma}H-G\frac{\partial H}{\partial \gamma}}}{H^2},
} \\
  {\displaystyle G=-k\cdot z,\;  H=(\alpha{V}+1)^2+
  {z}^2,
}  \\
 {\displaystyle \frac{\partial G}{\partial \gamma}=-\frac{\partial k}{\partial \gamma}\cdot z, \frac{\partial H}{\partial \gamma}=2(\alpha V+1)\alpha\frac{\partial V}{\partial \gamma}.
}
  \end{array}
\end{equation}
The first derivatives of functions $U$ and $k$ with respect to variables $\gamma$ and $\Delta$ in the above equation are the same as determined in Eqs (\ref{eq:dU-gamma}), (\ref{eq:dk-gamma}) and (\ref{eq:dk-d}).

Using the derivatives of the kernel obtained above, we can gain the second derivatives of Grover's formula of mutual inductance with respect to the appropriate coordinates and write  corresponding formulas for the calculation of magnetic stiffness. Taking into account Eqs (\ref{eq:second rho-der of kernel Grover}), (\ref{eq:second z-der of kernel Grover}) and (\ref{eq:second drho-der of kernel Grover}), we can write
\begin{equation}\label{eq:Srho GF}
\begin{array}{l}
  {\displaystyle S_{\rho\rho}=-\frac{\mu_0I_pI_s}{2\pi}\frac{\sqrt{R_p}}{R_s^{1.5}}\int_{0}^{2\pi}\frac{\partial^2 U}{\partial  \gamma^2} \cdot\Psi(k)+2\frac{\partial U}{\partial  \gamma}\cdot\frac{d \Psi(k)}{d k} \cdot\frac{\partial k}{\partial  \gamma} }\\
   {\displaystyle  +U\left(\frac{d^2 \Psi(k)}{d k^2} \cdot\left(\frac{\partial k}{\partial  \gamma}\right)^2+\frac{d \Psi(k)}{d k} \cdot\frac{\partial^2 k}{\partial  \gamma^2} \right)\;\;d\varphi,}
\end{array}
\end{equation}
\begin{equation}\label{eq:Sd GF}
\begin{array}{l}
  {\displaystyle S_{dd}=-\frac{\mu_0I_pI_s}{2\pi}\frac{\sqrt{R_s}}{R_p^{1.5}}\int_{0}^{2\pi} U\left[\frac{d^2 \Psi(k)}{d k^2}\left(\frac{\partial k}{\partial \Delta}\right)^2+\frac{d \Psi(k)}{d k} \cdot\frac{\partial^2 k}{\partial \Delta^2} \right]\;d\varphi, }
\end{array}
\end{equation}
\begin{equation}\label{eq:Sdrho GF}
\begin{array}{l}
  {\displaystyle S_{\rho d}= S_{d\rho}=-\frac{\mu_0}{2\pi}\frac{I_pI_s}{\sqrt{R_pR_s}}\int_{0}^{2\pi}\frac{\partial U}{\partial  \gamma } \cdot\frac{d \Psi(k)}{d k}\cdot\frac{\partial k}{\partial  \Delta} }\\
   {\displaystyle +U\left(\frac{d^2 \Psi(k)}{d k^2} \cdot\frac{\partial k}{\partial  \gamma}\cdot\frac{\partial k}{\partial  \Delta}+\frac{d \Psi(k)}{d k} \cdot\frac{\partial^2 k}{\partial  \gamma \partial \Delta} \right)\;\;d\varphi.}
\end{array}
\end{equation}

\bibliography{References}

\end{document}